\documentclass[aps,showpacs,floatfix,preprint]{revtex4}

\usepackage{amsfonts}
\usepackage{amsmath}
\usepackage{amssymb}
\usepackage{amsthm}
\usepackage{epsf}
\usepackage[dvips]{graphicx}

\def\bk{{\mbox{\boldmath$k$}}}

\def\bp{{\mbox{\boldmath$p$}}}
\def\bgam{{\mbox{\boldmath$\gamma$}}}
\def\ds{\displaystyle}
\def\bxi{\mbox{\boldmath{$\xi$}}}
\begin{document}

\title {\Large \bf Two-Fermion
Bound States within the Bethe-Salpeter Approach}

\author {S.M. Dorkin} \email {dorkin@theor.jinr.ru}
\affiliation {International University Dubna, Dubna, Russia}
\author{M. Beyer}\email{michael.beyer@uni-rostock.de}
\affiliation{Institute of Physics University of Rostock, D-18051 Rostock, Germany}
\author{ S.S. Semikh}\email {semikh@theor.jinr.ru}\author {L.P. Kaptari}\email {kaptari@theor.jinr.ru}
\affiliation{Bogoliubov Lab. Theor. Phys. Joint Institute for Nuclear Research,141980
 Dubna, Moscow reg., Russia}

\begin{abstract}
To solve the spinor-spinor  Bethe-Salpeter equation
in   Euclidean space we propose a novel method related
to the use of hyperspherical harmonics.  We suggest an
appropriate extension to form a new basis of spin-angular
harmonics that is suitable for a representation of the  vertex functions.
We present a numerical algorithm to  solve  the Bethe-Salpeter equation
and investigate in detail the properties of the solution for the
scalar, pseudoscalar and vector  meson exchange kernels including
the stability of bound states. We also compare our results to the
non relativistic ones and to the results given by
light front dynamics.
\end{abstract}


\maketitle

\section{Introduction}
Interpretation of many modern experiments
requires a covariant description of the two-body system. This is
either due to high precision that calls for an inclusion of all possible
corrections to a standard (possibly nonrelativistic) approach or due
to the high energies and momenta involved in the processes investigated.
Even the simplest atomic object, the hydrogen atom, or the simplest leptonic
system, the Positronium, needs a covariant description to match the high
experimental precision already achieved. In the subatomic field the most obvious
examples are the properties and structure of the deuteron
and to some extent the mesons, if the later are  treated as a quark antiquark
system.
In the spirit of a local quantum field theory the starting point of a relativistic
covariant description of  bound states of two particles is the
Bethe-Salpeter (BS)  equation. However, despite the obvious simplicity of
 two-body systems, the procedure of solving the BS equation
 encounters difficulties. These are related to singularities
 and branch points (cuts) of the  amplitude
 along the  real axis of the relative
 energy in  Minkowski space. Therefore, up to now the BS equation
including realistic
  interaction kernels has been solved either in Euclidean space
 within the ladder approximation (see e.g. Refs.~\cite{rupp,tjon,ztjon,umnikov,maris}
 and references  quoted therein)
 or utilizing additional approximations
 of the equation itself~\cite{gross,LCTA,karmanov,tobias,efimov,bakker}.
 Doing so, a fairly good  description of experimental
 data has been achieved (cf. Refs. \cite{gross1}-\cite{our_brkp}).
 Note, however, that for processes involving the deuteron the  new  data
 on  electromagnetic form factors \cite{edff} and  on the
 proton-deuteron scattering reactions \cite{cosy}  are still challenging.
 In addition, there are open issues describing
the positronium as a bound  electron-positron state
interacting via electromagnetic forces~\cite{positr}.

 At  high energies a
 consistent relativistic analysis of processes with two particles
is of a great  importance since covariance and relativistic invariance
play a crucial role in the  calculation of  matrix elements.
At low energies a relativistic framework
can  also be  relevant, because of the high precision achieved
by modern experiments. This means that  a relativistic framework
might be  essential to
facilitate understanding of the properties of bound system even  at
non relativistic energies. E.g., investigating
the energy spectrum of the hydrogen atom as a bound system
of an electron and a proton that interact via the Coulomb
potential requires a refined description of the fine structure
of the spectrum that can be accomplished by using the relativistic Dirac equation
for the electron wave function~\cite{LL}.
Otherwise, in a pure non relativistic framework
a sizable amount of corrections to  the potential must be considered.
A similar scenario appears in hadronic  physics when analyzing deuteron
reactions mentioned above, were the deuteron is treated within the impulse approximation.
It is common notion that a non relativistic approach fails to describe
processes at intermediate and high energies unless additional
(relativistic) corrections are taken into account. At the
same time, the use of the BS amplitude
 already in the ''impulse approximation''  provides
such corrections, e.g. the Lorentz boost effects, meson exchange
currents etc~ \cite{const}, in a natural way. Hence,
 analyzing different hadronic and
electromagnetic processes involving light nuclei
\cite{static,dp,our_chex,our_brkp,const,edenp} we conclude
that the relativistic description of the initial and final states
of the reacting  particles should be included {\it ab initio}.

A realistic implementation of this program started
off by the pioneering works by J. Tjon~\cite{tjon}-\cite{ztjon}.
 The basic challenge  of the BS approach
and  that of  other relativistic formalisms, such as
the  quasipotential approach or equations of relativistic
quantum mechanics within    Light Front  Dynamics,
is  the covariant  description of the
nucleon-nucleon interactions.
As new refined experimental data on  deuteron reactions
at intermediate and high  energies have become available
(cf. Refs. \cite{edff,cosy,t20exp}), the interest in a theoretical
treatment of  relativistic equations has been renewed: the
procedure of solving
the BS equation  numerically has been revisited  in Ref.~\cite{fb-tjon},
a reduction of the BS equation to an equation of
the Light Front form has been
 proposed in Ref.~\cite{tobias}. Also, a detailed investigation of
the  solution and the properties of  two spinor
particle bound state within  Light Front Dynamics  has been reported
in Refs.~\cite{karm2,gsalme}.

Unfortunately, our understanding of the mathematical properties of bound
states within a relativistic approach is far from being perfect.
In mathematical terms the BS equation itself is a quite complicated object,
and the technical problem of solving it is still a fundamental
issue. Consequently there are very few successful
examples of solving the BS equation for fermions including realistic
interactions. For
example, in Refs.~\cite{tjon,umnikov} the BS equation for spinor particles was
regularly treated by using a two-dimensional Gaussian mesh. That
series of studies revealed a high feasibility of the BS approach to
describe nucleon-nucleon interactions, in particular, processes
involving the  deuteron.
However, it should be mentioned that the algorithms exploiting the two-dimensional
meshes are rather cumbersome and require large computer resources.
In addition, the numerical solution is obtained as two-dimensional
arrays  which are quite awkward in practice when computing matrix elements
and in attempts to establish  reliable parameterizations and
possible analytical continuations of the solution back to  Minkowski space.
 Therefore, it
is necessary to provide  a method for solving BS equations
that would feature a smaller degree of arbitrariness.

In the present paper we suggest an
efficient and promising method to solve the BS equation for
fermions involving interaction kernels of one-boson
exchange type supplemented by corresponding form factors. It is
based on hyperspherical harmonics used to
expand  partial amplitudes and kernels. We show that this
novel technique provides many insights into the BS
approach. The current study is partially stimulated
by the results
reported in Ref.~\cite{karmanov}. We explore the structure of $^1S_0$ and
$^3S_1 - ^3D_1$ bound states for different couplings and study the  details of
the convergence of solutions and corresponding eigenvalues. In
particular, on the basis of the proposed   method for solving
the BS equation it becomes possible to analyze the specifics
 of the problem related to the stability of bound states in the BS approach.
Besides that, the hyperspherical expansion provides an
effective parameterization of the  amplitude, which is extremely
useful in practical calculations of observables
and in theoretical investigations  of  the separability
of the BS kernel with
one-boson exchange interaction~\cite{last_yaf,arhiv}.

Our  paper is organized as follows.
Section II contains, in the context of scalar particles, an
overview on the use of the  hyperspherical harmonics
 which displays the basic features of the method.
 In Section III, our generalization of the method
 for the spinor-spinor BS equation is presented. We introduce
a new basis of spin-angular harmonics in the spinor space and present
the appropriate decomposition for the BS amplitude.  The corresponding
analytical expressions for the partial amplitudes are found explicitly
in  Euclidean space.
Numerical calculations together with an analysis of
the stability of bound states are presented in Section IV.
In this Section the computational algorithm is introduced
 and the results for scalar, pseudoscalar and
vector meson exchange  kernels
are discussed. We give our conclusions in Section V.
The most cumbersome formulas are collected in the Appendix.

\section{Overview of the method}

The main idea of solving multidimensional integral equations
such as the Bethe--Salpeter equation
consists in finding an appropriate decomposition of the unknown solution over
a complete basis in  momentum space, carry out several
integrations over this basis analytically and solve the resulting equations with respect to the
unknown coefficients of such a decomposition. In this way one reduces the dimension
of the relevant integrals and essentially simplifies the numerical procedure of
solving the equation. Usually (see e.g. \cite{tjon,umnikov}), the corresponding basis
is chosen to be the complete set of the two-dimensional spherical harmonics $Y_{lm}(\theta,\phi)$
which allows one to
eliminate all the angular dependencies from the corresponding equation. Since the BS equation
is a four-dimensional integral equation,  the decomposition over the spherical harmonics
 results in  an equation with only  two-dimensional integrations that has
to be  solved numerically. Instead, in the present paper we suggest a
 higher dimensional basis, i.e. the basis of hyperspherical  harmonics.  Apart from
the familiar spherical angles $\theta,\phi$, it also includes  a third variable -
the hyperangle $\chi$ (see, below). With this basis one can eliminate
up to three variables from
the four-dimensional integral equation and to reduce the problem
to find numerical solutions of ordinary (one-dimensional)   integral   equations.
To demonstrate the essence of the method, let us consider the simplest
case of the homogenous Bethe--Salpeter equation for
 two  scalar particles  with equal  masses $m$
interacting via exchange of a scalar  particle with
mass $\mu$~\cite{fb-tjon}. The corresponding BS equation for
 the vertex function ${\cal G}(p)$
is
\begin{eqnarray}
\label{sceq}
{\cal G}(p)=i\,g^2\int\frac{d^4k}{(2\pi)^4}\,V(p,k)\,S(k_1)S(k_2)\,{\cal G}(k),
\end{eqnarray}
where $k_{1,2}=P/2\pm k$, and $k=(k_0,\bk), p=(p_0,\bp)$ are the relative 4-momenta,
$P=(M,{\bf 0})$ is the total 4-momentum of the two particles
in their center of mass system and $g$ is the corresponding  coupling
constant. In (\ref{sceq})
$S(k_{1,2})$ and $V(p,k)$ are the free propagators of the
constituents and of the exchanged particle, respectively
\begin{eqnarray}
S(k_{1,2})=\frac{1}{k_{1,2}^2-m^2+i\varepsilon},\quad
V(p,k)=\frac{1}{(p-k)^2-\mu^2+i\varepsilon}\label{vpk}.
\end{eqnarray}
It is seen from  (\ref{vpk}) that even  in this simplest scalar
 case the BS equation
contains singularities (poles) in   Minkowski space.
Moreover, it is known
that the vertex function ${\cal G }(p)$ itself contains  cuts along the real axis
of $p_0$ ($k_0$)
making the solving procedure rather cumbersome.
To  rid us of difficulties connected to the treatment of these singularities,
one usually performs the Wick rotation in the complex plane $(p_0,ip_4)$   and solves the
BS equation  (\ref{sceq}) in  Euclidean space, where it
is given by
\begin{eqnarray}
{\cal G}(p_4,\bp)=g^2 \int \frac {d^4k}{(2\pi)^4} \frac {1}{(p-k)^2+\mu^2}
\frac {{\cal G}(k_4,\bk)}{(k^2+m^2-M^2/4)^2+M^2k_4^2}.
\label{sceqe}
\end{eqnarray}
To reduce the dimension of the integral
 we decompose the relevant quantities in (\ref{sceqe})  over the basis of
 hyperspherical harmonics $Z_{nlm}(\chi,\theta,\phi)$

\begin{eqnarray}
Z_{nlm}(\chi,\theta,\phi)&=&X_{nl} (\chi) Y_{lm}(\theta,\phi),\nonumber\\
X_{nl}(\chi)&=&\sqrt{\frac{2^{2l+1}}{\pi} \frac
{(n+1)(n-l)!l!^2}{(n+l+1)!}} \sin^l\chi C_{n-l}^{l+1}(\cos \chi),
\label{hsg}
\end{eqnarray}
where $Y_{lm}(\theta,\phi)$ are the familiar spherical harmonics,
and $C_{n-l}^{l+1}$ are
the Gegenbauer polynomials.
The hyperangle $\chi$ of a 4-vector $p=(ip_4,{\bf p})$  is defined as
\begin{equation}
\cos \chi =\frac{p_4}{\tilde p};\qquad
\sin\chi =\frac{|{\bf p}|}{\tilde p}
\label{gspha}
\end{equation}
with Euclidean 4-vectors $p$ and $k$
having modules $\tilde p=\sqrt{p_4^2+\bp^2}$ and $\tilde k=\sqrt{k_4^2+\bk^2}$.
The hyperspherical harmonics
  (\ref{hsg})   satisfy the orthonormalization
relation
\begin{eqnarray}
\int_0^{2\pi} d\phi \int _{0}^{\pi} d\theta \sin \theta
\int_{0}^{\pi} d\chi \sin^2 \chi Z_{klm}(\chi, \theta, \phi)
Z^*_{k^{\prime}l^{\prime}m^{\prime}}(\chi, \theta,\phi)=\delta_{kk^{\prime}} \delta_{ll^{\prime}}
\delta_{mm^{\prime}}.
 \label{orto}
\end{eqnarray}
For the interacting kernel in (\ref{sceqe}) it follows
\begin{eqnarray}
\frac {1}{(p-k)^2+\mu^2}&=&2\pi^2 \sum_{nlm}
\frac{1}{n+1} V_n(\tilde p,\tilde k)Z_{nlm}(\chi_p, \theta_p,\phi_p)
Z_{nlm}^{*}(\chi_k, \theta_k,\phi_k),\label{exp12}
\\
V_n(\tilde p,\tilde k)&=& \frac
{4}{(\Lambda_+ +\Lambda_-)^2}
\left( \frac {\Lambda_+ -\Lambda_-}{\Lambda_+ +\Lambda_-}\right)^n,
\label{vn} \\
\Lambda_\pm&=&\sqrt{(\tilde p \pm \tilde k)^2+\mu^2}.
\nonumber
\end{eqnarray}
The vertex function is then given in the form
\begin{eqnarray}
\label{exp11}
{\cal G} (k_4,\bk)&=&\sum\limits_{nlm}\varphi_l^n
(\tilde k)\,Z_{nlm}(\chi_k,\theta_k,\phi_k).
\label{er}
\end{eqnarray}
Changing the integration variables,
$d^4k =\tilde k^3\sin^2\chi_k \sin\theta_k d \tilde k d\chi_k  d\theta_k d\phi_k$,
inserting (\ref{exp12}) and (\ref{er}) into
 (\ref{sceqe}) and performing the necessary angular integrations we
obtain  a system of one-dimensional integral equations for
the expansion coefficients~$\varphi_l^n$
\begin{eqnarray}
\varphi_l^n(\tilde p) =
\int \frac {d\tilde k \tilde{k}^3}{8\pi^2}
V_{l}(\tilde p, \tilde k) \sum \limits_{m=1}^{\infty}
A^{nm}_l(\tilde k) \varphi_l^m(\tilde k)
\label{eqnf}.
\end{eqnarray}
The explicit expressions for the coefficients $A^{nm}_l(\tilde k) $
result from the corresponding angular integrations.
Note that  a  two
particle bound state is characterized by definite values of angular momenta $l$,
so that only few (one) values of $l$ contribute  in  (\ref{er}) and (\ref{eqnf}).

Equation (\ref{eqnf}) demonstrates how  one can obtain an equivalent
system of one dimensional integral
equations by starting from the four-dimensional integral
 equation (\ref{sceqe}) and applying   the hyperspherical harmonics decomposition
 (\ref{er}). Formally the expansion that leads to the
final equation (\ref{eqnf})  contains an infinite number of terms,
$(n,m=1,\infty)$,
hence, from a pure mathematical point of view,
the problem of finding a solution seems not much simplified.
However, in practise when starting from an approximate solution of
(\ref{sceqe}), it turns out that the series (\ref{er}) converges
rather fast and it suffice to
restrict oneself to a finite number of terms,
  making then the system finite and mathematically meaningful.
Then the procedure of solving  the system
 (\ref{eqnf}) numerically becomes rather straightforward.
 Previously we found  that the first
 three to four terms in the decomposition (\ref{er})
 assure a sufficiently  high accuracy of the solution \cite{last_yaf,arhiv}.
 Only in the case of a low binding energy $M$ and
 very light exchanged particles, $\mu\to 0$,
 more  terms are needed in the series (\ref{er}) for a convergence.
 In the limit  $\mu=0$ a more extended analysis is required.
Note that the described method can also be applied to the inhomogeneous
BS equation.

\section{Spinor-Spinor  BS equation }
Below we generalize the method  introduced in the previous section to
the spinor-spinor BS equation,
i.e., the BS equation for two spinor particles interacting via
one-boson-exchange potentials. In this case the   vertex
${\cal G}(p)$ is a
$4\times 4$ matrix in  spinor space and the
corresponding BS equation is~\cite{nakan}
\begin{eqnarray}
\label{sphom}
{\cal G}(p)=i g^2 \int\frac{d^4k}{(2\pi)^4}\,V(p,k)\,\Gamma(1)\, S(k_1)\,{\cal
G}(k)\,{\tilde S(k_2)}\,\tilde\Gamma(2),
\end{eqnarray}
 where  the propagator $V(p,k)$
 for the  exchange of scalar and pseudoscalar mesons is given in (\ref{vpk})
and by
\begin{eqnarray}
V(p,k)=\frac{-g_{\mu\nu}+\ds\frac{q_\mu q_\nu}{\mu^2}}{q^2-\mu^2+i\varepsilon}
\label{vecprop}
\end{eqnarray}
 for vector mesons.
The   propagators of the spinor  constituents
  are
\begin{eqnarray}
\nonumber {S}(k_1)=
\frac{\hat k_1 +m}{k_1^2-m^2+i\varepsilon},\quad
{\tilde S}(k_2)\equiv C{S}(k_2)^TC=\frac{\hat k_2
-m}{k_2^2-m^2+i\varepsilon},
\end{eqnarray}
with  the charge conjugation matrix
$C=i\gamma^0\gamma^2$. The meson vertices $\Gamma$
are determined by the corresponding effective interaction Lagrangians
describing the interaction of the spinor particles with mesons. For a system
of two nucleons
  they  are given by
\begin{align}
\Gamma(1)&= 1&\tilde\Gamma(2)&=- 1&\text{scalar}\\
\Gamma(1)&= \gamma_5&\tilde\Gamma(2)&=-\gamma_5&\text{pseudoscalar}\\
\Gamma(1)&=\gamma_\mu-\frac{i\kappa}{2m}\,\sigma_{\mu\rho}q^\rho&
\tilde\Gamma(2)&=\gamma_\nu+\frac{i\kappa}{2m}\,\sigma_{\nu\rho}q^\rho&\text{vector}
\label{v3}
\end{align}
In (\ref{v3}) the momentum transfer $q$ is defined as $q=p-k$,
$\kappa$ denotes the strength of the tensor part of the interaction and the coupling
constant $g$ in (\ref{sphom}) is imaginary  for the pseudoscalar mesons else real.
Each interaction vertex $\Gamma$ is regularized by a monopole form factor
\begin{eqnarray}
F(q^2)=\frac {\Lambda^2-\mu^2}{\Lambda^2-q^2}
\label{ff}
\end{eqnarray} with $\Lambda$ as free  cut-off parameters.
For the sake of simplicity we presently adopt $\kappa=0$.
Consequently, in the propagators of vector particles,
eq.  (\ref{vecprop}), the part  proportional to
$q_\mu q_\nu / \mu^2$  is neglected as well.
 These restrictions lead only to
slight redefinitions of other effective constants  (cut-off parameters,
vector coupling constants etc) and do not affect the method and the main
final conclusions.

Contrary to the scalar case, the bound state of two spinor particles
is characterized by the total angular momentum $J$, which is an algebraic
 sum of the total spin $S=0,1$ and total orbital momentum $L$
($L=0,1,2....$), i.e., $J=S+L$.
Traditionally, for  two spinor  bound states one
adopts the spectroscopic notation $^{2S+1}L_J$.

 We investigate the  lowest ground states of the
 $^1S_0$ and $^3S_1$--$^3D_1$ channels taking into account
 only one type of mesons at a time, either
 scalar, pseudoscalar or vector. The inclusion of the sum of all mesons
in the  interaction kernels of the BS equation which reflects
the more realistic case is beyond
the scope of the present paper. A
 generalization of the method will be done elsewhere.

For completeness we give the  BS  amplitude  $\Psi$ used in the following
that is related to the
corresponding vertex functions via
\begin{eqnarray}
\Psi(p)=S(p_1){\cal G}(p) \tilde{S}(p_2).
\label{BSA}
\end{eqnarray}

\subsection{Spin-angular harmonics}
The main difference between the scalar  (\ref{sceqe}) and the
spinor BS equation (\ref{sphom}) is that in the later case
the vertex function  ${\cal G}(p)$  is a
$4\times 4$ matrix in  spinor space. Consequently, the
spinor BS equation is of
matrix form and before proceeding with numerical solutions we shall
transform it into a system of ordinary equations. To this end,
we expand    ${\cal G}(p)$ into a complete
set of $4\times 4$ basis  matrices and obtain
a system of equations for the coefficients of such a decomposition. In the most general case,
there are 16 linearly independent   matrices that can be used as a basis.
The  choice of such a basis system depends on
the specific aim envisaged. Usually one uses
either the complete set of the Dirac matrices \cite{umnikov} or
the complete set of the spin-angular harmonics \cite{kubis}.
Different basis are related to each other via unitary
transformations \cite{static}.

For specific bound states with given quantum numbers
only some basis matrices contribute
to the vertex function  ${\cal G}(p)$ (amplitude $\Psi(p)$  ).
E.g., for the
$^1S_0$ state only four matrices
are relevant to describe the amplitude, while
in the $^3S_1-^3D_1$ channel eight basis matrices are needed.
 A  standard choice for the basis matrices in these two channels
is referred to as $\rho$-spin angular harmonics
$\Gamma_\alpha(\bp)$, where the index $\alpha$
 lists the quantum numbers of the  $LSJ$ momenta
 and that of the $\rho$--spin. In fact, the spin angular harmonics
 are constructed as outer products of two Dirac spinors
 which, for each constituent, form  complete sets of
  solutions of the free Dirac equation.
  Then the $\rho$-spin (projection) reflects the signs of the energy of two basis
   spinors in the outer  product and is labeled by
 $++$,$--$,$+-$ and $-+$ correspondingly. Sometimes, instead of
$+-$ and $-+$ one uses linear combinations which can be even
($e$) or odd ($o$) with respect to symmetry on the relative energy $p_0$
 (for details see Ref.~\cite{kubis}).
In Table~\ref{tab1}
we present   the classification of the partial components
in this basis and the spectroscopical notation for the partial components of the BS
amplitude in the $^1S_0$ channel (first row) and in
the $^3S_1$--$^3D_1$ channel (second row). The explicit expressions
for the corresponding spin angular harmonics $\Gamma_{\alpha}(\bp)$
can be found, e.g., in Ref.~\cite{static}.

The expansion of the BS
amplitude $\Psi(p_0,\bp)$ and of the vertex function ${\cal G}(p_0,\bp)$
into spin-angular harmonics is
\begin{eqnarray}
\Psi(p_0,\bp)&=&\sum\limits_\alpha \psi_\alpha(p_0,|\bp|)
\,\Gamma_\alpha(\bp),\label{spex1} \\
{\cal G}(p_0,\bp)&=&\sum\limits_\alpha g_\alpha(p_0,|\bp|)
\,\Gamma_\alpha(-\bp)
\label{spex},
\end{eqnarray}
where the expansion coefficients $\psi_\alpha(p_0,|\bp|)$ and
$g_\alpha(p_0,|\bp|)$ are evaluated numerically.
A comparison of (\ref{spex}) with (\ref{hsg}) and (\ref{er})
shows that the spin-angular harmonics $\Gamma_\alpha$
can be considered as a direct generalization of the spherical harmonics
$Y_{lm}$ in spinor space. Hence, at first glance, it seems  natural
to  expand also  the
coefficients $g_\alpha(p_0,|\bp|)$ ($\psi_\alpha(p_0,|\bp|)$)
into Gegenbauer polynomials $X_{nl}(\chi)$ defined in (\ref{hsg})
and to obtain a system of one dimensional integral equations
like (\ref{eqnf}). However, a more detailed inspection of the spinor-spinor
BS equation, together with the explicit forms of the spin-angular harmonics
shows that the use of the commonly accepted form of the
$\rho$-spin basis hinders a further use of the Gegenbauer
polynomials. To be more specific, note that in  (\ref{spex1}) and (\ref{spex})
terms appear that  always mix the angular and hyperangular
dependence (viz. nonlinear terms proportional to, e.g. $1/|\bp|$) which  make an
analytical integration over
the hyperangle $\chi$ impossible.
To avoid such difficulties within the hyperspherical harmonics formalism  we suggest
a slightly modified  set of spin-angular matrices which
represent a generalization of the
$\rho$-spin basis.
In the $^1S_0$ channel  the new basis is
\begin{eqnarray}
{\cal T}_1(\bp)&=&\frac{1}{\sqrt{16\pi}}\gamma_5,\nonumber\\
{\cal T}_2(\bp)&=&\frac{1}{\sqrt{16\pi}}\gamma_0\gamma_5,\nonumber \\
{\cal T}_3(\bp)&=&\frac{1}{\sqrt{16\pi}}\frac{(\bp \bgam)}{|\bp|}\gamma_0\gamma_5,\nonumber \\
{\cal T}_4(\bp)&=&\frac{1}{\sqrt{16\pi}}\frac{(\bp\bgam)}{|\bp|}\gamma_5,
\label{nharms}
\end{eqnarray}

and for the $^3S_1$--$^3D_1$ channel we choose
\begin{eqnarray}
\nonumber
{\cal T}_1(\bp)&=&-\frac{1}{\sqrt{16\pi}}(\bgam \bxi_{\cal M}),\\
\nonumber
{\cal T}_2(\bp)&=&-\frac{1}{\sqrt{16\pi}}\,\gamma_0\,(\bgam \bxi_{\cal M}),\\
\nonumber
{\cal T}_3(\bp)&=&\frac{\sqrt{3}}{\sqrt{16\pi}}\frac{(\bp \bxi_{\cal M})}{|\bp|},\\
\nonumber
{\cal T}_4(\bp)&=&\sqrt{\frac{3}{2}}\frac{1}{\sqrt{16\pi}}\,\frac{\gamma_0}{|\bp|}\left[(\bp \bxi_{\cal M})+
(\bp \bgam)\,(\bgam \bxi_{\cal M})\right],\\
\label{nharmd}
{\cal T}_5(\bp)&=&\frac{\sqrt{2}}{2}\frac{1}{\sqrt{16\pi}}\left[-(\bgam \bxi_{\cal M})+\frac{3}{|\bp|^2}\,
(\bp \bxi_{\cal M})(\bp \bgam)\right],\\
\nonumber
{\cal T}_6(\bp)&=&\frac{\sqrt{2}}{2}\frac{1}{\sqrt{16\pi}}\gamma_0
\left[-(\bgam \bxi_{\cal M})+\frac{3}{|\bp|^2}\,
(\bp \bxi_{\cal M})(\bp \bgam)\right],\\
\nonumber
{\cal T}_7(\bp)&=& \frac{\sqrt{3}}{\sqrt{16\pi}}\,\gamma_0\frac{(\bp \bxi_{\cal M})}{|\bp|},\\
\nonumber
{\cal T}_8(\bp)&=&\sqrt{\frac{3}{2}}\frac{1}{\sqrt{16\pi}}\,\frac{1}{|\bp|}\left[(\bp \bxi_{\cal M})+
(\bp \bgam)\,(\bgam \bxi_{\cal M})\right].
\end{eqnarray}
The left hand side of (\ref{nharmd}) depends implicitly on $\mathcal{M}$, which denote the components
of the polarization vector $\bxi_{\cal M}$ fixed by
$\bxi_{+ 1} =-(1,i,0)/\sqrt{2}$, $\bxi_{- 1} =(1,-i,0)/\sqrt{2}$,
$\bxi_{0} =(0,0,1)$.
The new basis is   orthogonal and normalized
\begin{eqnarray}
\int d\Omega_p\, \mathrm{Tr}\,[{\cal T}_{m\mathcal{M}}(\bp)
{\cal T}_{n\mathcal{M}'}^+(\bp)]=\delta_{mn}\delta_{\cal MM'}
\nonumber.
\end{eqnarray}

The partial
expansions of the vertex functions over   the new basis is given by
\begin{eqnarray}
{\cal G}(p_0,\bp)=\sum\limits_{n}g_n(p_0,|\bp|) \,{\cal T}_n(-\bp)\label{exp1},
\end{eqnarray}
with
\begin{eqnarray}
g_n(p_0,|\bp|)= \int d\Omega_p\, \mathrm{Tr}\,[{\cal G}(p_0,\bp){\cal
T}_n^+(-\bp)] \label{exp2}.
\end{eqnarray}
In (\ref{exp1}) $n=1\dots n_\mathrm{max}$, where $n_\mathrm{max}=4$ for the $^1S_0$ and
$n_\mathrm{max}=8$ for the  $^3S_1-^3D_1$
bound states. As mentioned, different
bases are related to each other via corresponding unitary transformations.
The connection between the $\rho$-spin  and the new basis (\ref{nharmd}),
 or equivalently,
the relation between the quantities $g_\alpha$ ($\alpha=JLS\rho$)
and $g_n$ ($n=1\ldots n_{max}$)
can be found by using the completeness of the two basis and the parity of
the coefficients $g_i$ with respect to the operation
$ {p_0}\to -p_0$.
 The Pauli principle together with charge conjugation operation   $C$
 leads to
\begin{eqnarray}
C{\cal T}_n^T(-\bp)C=\pi_{p_0}\,(-1)^{I-1}{\cal T}_n(\bp),
\label{pau}
\end{eqnarray}
where $I$ stands for the isospin of the system, $\pi_{p_0}$ is the $\rho$-spin
parity.
From (\ref{pau}) one  obtains that
in the $^1S_0$ channel the component $g_4$ is of the odd  parity ($\pi_{p_0}=-1$)
while $g_1, ..., g_3$ are of the even parity ($\pi_{p_0}=+1$); correspondingly, in
the $^3S_1- ^3D_1$ channel the two components $g_7, g_8$ are odd in $p_0$,
and $g_1, ..., g_6$ are even.

The unitary relation between two sets eq. (\ref{spex}) and
 eqs. (\ref{nharms}) and (\ref{nharmd}) is
\begin{eqnarray}
g_{^1S_0^{++}}&=&\frac {1}{\sqrt 2} g_1+ \frac {1}{\sqrt 2}\frac{m}{E_p} g_2-\frac {1}{\sqrt 2}\frac{|\bp|}{E_p}g_3,
 \nonumber \\
g_{^1S_0^{--}}&=&-\frac {1}{\sqrt 2} g_1+ \frac {1}{\sqrt 2}\frac{m}{E_p}g_2-\frac {1}{\sqrt 2}\frac{|\bp|}{E_p}g_3,
\label{1s0} \\
g_{^3P_1^o}&=& -\frac {|\bp|}{E_p}g_2 - \frac{m}{E_p}g_3,\quad
g_{^3P_1^e}=g_4,\nonumber
\end{eqnarray}
for the  $^1S_0$ states and
\begin{align}
\left(
\begin{array}{@{}c@{}}
g_{^3 S_1^{++}}\\
g_{^3 S_1^{--}}\\
g_{^3 D_1^{++}}\\
g_{^3 S_1^{--}}\\
g_{^1P_1^e} \\
g_{^3P_1^o}
\end{array}
\right )
&=
\left (
\begin{array}{@{}cccccccc@{}}
 \frac{\sqrt 2}{6}\frac{m+2E_p}{E_p}  &  \frac{\sqrt 2}{6} \frac{2m+E_p}{E_p}  &-
 \frac{1}{\sqrt 6}\frac {|\bp|}{E_p}  & - \frac{1}{\sqrt 3} \frac{|\bp|}{E_p}  &
  \frac {1}{3} \frac {E_p-m}{E_p}     & -\frac {1}{3} \frac {E_p-m}{E_p}  \\
- \frac{\sqrt 2}{6}\frac{m+2E_p}{E_p} & \frac{\sqrt 2}{6} \frac{2m+E_p}{E_p} &-
 \frac{1}{\sqrt 6}\frac {|\bp|}{E_p}  & -\frac{1}{\sqrt 3} \frac{|\bp|}{E_p}  &
-\frac {1}{3} \frac {E_p-m}{E_p}      & -\frac {1}{3} \frac {E_p-m}{E_p}      \\
-\frac{1}{3} \frac{E_p-m}{E_p}        & \frac{1}{3}\frac{E_p-m}{E_p}          &-
 \frac{1}{\sqrt 3}\frac{|\bp|}{E_p}   &-\frac{1}{\sqrt 6} \frac {|\bp|}{E_p} &
-\frac {\sqrt 2}{6} \frac {2m+E_p}{E_p}& -\frac {\sqrt 2}{6} \frac {m+2E_p}{E_p}  \\
 \frac{1}{3} \frac{E_p-m}{E_p}        &  \frac{1}{3}\frac{E_p-m}{E_p}   &
  \frac{1}{\sqrt 3}\frac{|\bp|}{E_p}  &\frac{1}{\sqrt 6} \frac {|\bp|}{E_p}&
\frac {\sqrt 2}{6} \frac {2m+E_p}{E_p}& -\frac {\sqrt 2}{6} \frac {m+2E_p}{E_p}  \\
\frac {1}{\sqrt 3}\frac {|\bp|}{E_p}  & 0&+\frac{m}{E_p} & 0&
-\frac {\sqrt 6}{3} \frac {|\bp|}{E_p}& 0\\
  0& \frac {\sqrt 6}{ 3}\frac {|\bp|}{E_p}& 0& \frac{m}{E_p}&0& \frac {1}{\sqrt 3} \frac {|\bp|}{E_p}
\end{array}
\right)
\left(
\begin{array}{@{}c@{}}
g_1\\
g_2\\
g_3\\
g_4\\
g_5 \\
g_6 \\
\end{array}
\right )
\nonumber\\[2mm]
g_{^1P_1^0}&=g_7, \quad g_{^3P_1^e}=g_8\label{3s1}
\end{align}
for the   $^3S_1-^3D_1$ state. Here $E_p$ denotes the total energy of
a free particle with the momentum $\bp$, i.e.
 $E_p=\sqrt{\bp^2+m^2}$. As seen from
 (\ref{1s0}) and (\ref{3s1}) all   quantities containing
 $E_\bp$ and $|\bp|$  that prevent the use of
Gegenbauer polynomials  have been  explicitly extracted
into the  corresponding coefficients.

To complete this section we present the
 partial amplitudes in terms of the corresponding
 partial components of the vertex functions
 in the $\rho$-spin basis
\begin{eqnarray}
\psi_{S(D)^{++}}(p_0,|\bp|) &=&
\frac {g_{S(D)^{++}}(p_0,|\bp|)}{(M/2-E_p)^2-p_0^2}, \label{ppp}\\
\psi_{S(D)^{--}}(p_0,|\bp|) &=&
\frac {g_{S(D)^{--}}(p_0,|\bp|)}{(M/2+E_p)^2-p_0^2}, \\
\psi_{P_e}(p_0,|\bp|) &=& \frac
{g_{P_e}(p_0,|\bp|)(M^2/4-p_0^2-E_p^2)+2g_{P_o}(p_0,|\bp|)p_0 E_p}
{(M^2/4-p_0^2-E_p^2)^2-4p_0^2E_p^2}, \\
\psi_{P_o}(p_0,|\bp|) &=& \frac
{g_{P_o}(p_0,|\bp|)(M^2/4-p_0^2-E_p^2)+2g_{P_e}(p_0,|\bp|)p_0 E_p}
{(M^2/4-p_0^2-E_p^2)^2-4p_0^2E_p^2}.
\end{eqnarray}
Since the components with negative $\rho$-spins reflect the contribution
 of  negative energies (of the solution of the Dirac equation)
one might argue that at moderate energies and momentum transfers they could
 be suppressed compared to the components with positive
 $\rho$-spins.  Hence in calculations of matrix elements
  such components could therefore safely be neglected. Moreover, often
 the components with mixed $\rho$-spins
 ($+-$ or $-+$) could also be disregarded in concrete calculations.
 Note, however, that in the BS equation  these components cannot be omitted,
 since the convergence and stability of the numerical solution is sensitive
 to each component.  To estimate the magnitude
of different  components one might introduce the  pseudo probability, i.e. the
contributions of each component in the  normalization condition
\begin{eqnarray}
P_{S(D)^{++}}&=&\int dp_0\,d|\bp|\,(E_p-M/2)\,|\psi_{S(D)^{++}}(p_0,|\bp|)|^2,\nonumber \\
P_{S(D)^{--}}&=&-\int dp_0\,d|\bp|\,(E_p+M/2)\,|\psi_{S(D)^{--}}(p_0,|\bp|)|^2,
\label{pseudopr} \\
P_e&=&-M \int dp_0\,d|\bp|\,|\psi_{P_e}(p_0,|\bp|)|^2,\nonumber\\
P_o&=&-M \int dp_0\,d|\bp|\,|\psi_{P_o}(p_0,|\bp|)|^2.\nonumber
\end{eqnarray}

\subsection{Hyperspherical decomposition}

  Similar to the scalar case  the spinor-spinor BS equation
  after the Wick rotation~\cite{wick},
  is considered
  in  Euclidean space (c.f. eq. (\ref{sceqe})).
  The Wick rotation can be achieved by replacing the
  scalar product of two vectors in  Minkowski space
  by their Euclidean analogue
  ($k^2=k_0^2-\bk^2\to  \tilde k^2=k_4^2+\bk^2$)
  and changing $p_0\to ip_4$ and
  $k_0\to ik_4$. Notice, that every odd function of $p_0$,
  being  homogeneous in its argument,
   explicitly receives  an additional imaginary unit
$i$ after the Wick rotation.
 Therefore, for convenience, we divide this common
 factor by redefining the odd partial components
\begin{eqnarray}
\nonumber g_4\to ig_4
\end{eqnarray}
for the $^1S_0$ channel and
\begin{eqnarray}
\nonumber g_7\to ig_7,\quad g_8\to ig_8.
\end{eqnarray}
for the  $^3S_1- ^3D_1$ channel.
Now, by placing eq. (\ref{exp1}) into the Wick rotated BS equation
(\ref{sphom}) and using (\ref{exp2}) one obtains
\begin{eqnarray}
g_n( p_4,|\bp|)= g^2 b_n \int d \Omega _{p}
\int\frac{d^4k}{(2\pi)^4} S(k_4, |\bk|)
\frac {1}{(p-k)^2+\mu^2}
\sum\limits_{m}A_{nm}(p,k)g_m(k_4,\bk)
\label{par1s0},
\end{eqnarray}
where
$m,n=1\ldots 4$ for the $^1S_0$  and
$m,n=1\ldots 8$ for the $^3S_1$--$^3D_1$ state, respectively, $b_n$ is a sign coefficient
reflecting the type of the exchanged meson (see Tables \ref{tab5} and \ref{tab6}) and
the  scalar part $S(k_4,|\bk|)$ of the two spinor propagators is defined as
\begin{eqnarray}
\label{prop}
S(k_4,|\bk|)=\frac{1}{\left(k^2+  m^2-\frac{M^2}{4}  \right)^2+M^2k_4^2}.
\end{eqnarray}
   The coefficients  $A_{nm}(p,k)$  in
(\ref{par1s0}) directly follow from
   calculating traces of the BS equation after
   multiplying it from the right  by the
   corresponding basis matrix (\ref{nharms}) or (\ref{nharmd}).
   The angular dependence of $A_{nm}(p,k)$  on  $\Omega_p$
   is entirely governed by the dependence on the vector
   $\bp$ of the  corresponding
   harmonics ${\cal T}_n(\bp)$~\footnote{in  one  keeps the transversal part in the
    meson propagator (\ref{vecprop})
    and in the tensor part of the  interaction vertex
    (\ref{v3}) the calculations of the angular dependence of $A_{nm}(p,k)$ become
    more cumbersome,   but however straightforward}. So, from
    (\ref{nharms}) one
   infers that in  the $^1S_0$ channel $A_{1m}$ and  $A_{2m}$ are $\propto Y_{00}(\hat\bp)$ while
   $A_{3m}$ and  $A_{4m}$  are $\propto Y_{1-\mu}(\hat\bk)  Y_{1\mu}(\hat\bp)$.
    Analogously from eq. (\ref{nharmd}) it can be found that
   in the $^3S_1-^3D_1$ channel the angular dependence of $A_{nm}(p,k)$
   is proportional to $ Y_{l\mu}(\hat\bp)$ with $l=0,1,2$.
    Such a  dependence
   essentially simplifies integrations over   $\Omega_p$.

After  expanding the interaction kernel  into hyperspherical
 harmonics, eq.  (\ref{exp12}),  integrations over the
 angles $\Omega_p$ and $\Omega_k$ are carried out analytically.
The result is
\begin{equation}
g_n( p_4,|\bp|)= g^2 b_n
\int\frac{k^3 dk \sin^2 \chi_k d\chi_k}{(4\pi^3)} S(k_4, |\bk|)
W_{l_n}(\tilde p,\tilde k,\chi_p,\chi_k)
\sum\limits_{m}a_{nm}(k_4,\bk)g_m(k_4,\bk)
\label{par2s0},
\end{equation}
where
\begin{eqnarray}
W_{l_n}(\tilde p,\tilde k,\chi_p,\chi_k) =
\sum_l \frac{1}{l+1}  V_l(\tilde p,\tilde k) X_{l l_n}(\chi_p) X_{l l_n}(\chi_k)
\nonumber
\end{eqnarray}
 is the hyperspherical partial kernel in Euclidean space.
The values of the angular momentum $l_n$ are restricted by the
dependence of $A_{nm}$ on $\Omega_p$ of the corresponding integrals.
These are
$l_n=0$ for $n=1,2$, and $l_n=1$
for $n=3,4$ in the $^1S_0$-channel
and  $l_n=0$
for $n=1,2$, $l_n=1$ for $n=3,4,7,8$ and $l_n=2$ for $n=5,6$
in the  $^3S_1-^3D_1$ channel
(see also eqs. (\ref{nharms}) and (\ref{nharmd})).
The explicit expressions for the quantities
$a_{nm}$ in eq. (\ref{par2s0}) are collected in Tables
\ref{tab2}, \ref{tab3} and \ref{tab4}. A prominent feature of
the  partial BS equation (\ref{par2s0}) that is related to
the proposed basis (\ref{nharms}) and (\ref{nharmd}) is that
the coefficients $a_{nm}$
are independent on the type of the exchanged meson in the interaction
kernel. This  dependence   is solely contained in the coefficients
$b_n$ (Tables~\ref{tab5} and \ref{tab6} )
and in the coupling constant $g$ (if a cut off form factor is
included into the consideration, the cut off parameter $\Lambda$ can also
depend on the meson type).
 From   Tables ~\ref{tab5}-\ref{tab4}
 one can also infer  the parity $\pi_{p_0}$ of the component
$g_n(p_4,|\bp|)$.
Note that in eq. (\ref{par2s0}) one can easily express
$S(k_4,|\bk|) $ of eq. (\ref{prop}) and $a_{nm}(k_4,|\bk|)$,
given in Tables \ref{tab2}, \ref{tab3} and \ref{tab4},
via the hyperspherical variables $\chi_k$ and $\tilde k$.

We now expand the partial vertex functions
into hyperspherical functions in a similar way
  done in eqs. (\ref{exp12}) and (\ref{exp11}).
Since
the value of the angular momentum  $l_n$ are
restricted in the  partial components $g_n$,
 as discussed before, the hyperspherical expansion
 reduces to an expansion into the functions
 $X_{jl}(\chi_p)$, i.e.  the Gegenbauer polynomials. Moreover,
  due to definite parity of the
components $g_n$ with respect to
 $\cos\chi_p$ ($\cos\chi_p\equiv p_4/\tilde p$),  summation over  $j$
is restricted to only even or only odd
values of  $j$, in accordance with the relation
\begin{eqnarray}
C_j^l(-\cos\chi_p)=(-1)^jC_j^l(\cos\chi_p).
\nonumber
\end{eqnarray}
For the partial BS components $g_n$ one gets
\begin{eqnarray}
g_{1,2}(p_4,|\bp|)&=&\sum_{j=1}^\infty
g_{1,2}^j(\tilde p)\,X_{2j-2,0}(\chi_p),
\label{s0p1}\\
g_{3}(p_4,|\bp|)&=&\sum_{j=1}^\infty
g_{3}^j(\tilde p)\,X_{2j-1,1}(\chi_p),\\
g_{4}(p_4,|\bp|)&=&\sum_{j=1}^\infty
g_{4}^j(\tilde p)\,X_{2j,1}(\chi_p).
\label{s0p2}
\end{eqnarray}
Similarly for the   $^3S_1$--$^3D_1$ channel we obtain
\begin{eqnarray}
g_{1,2}(p_4,|\bp|)&=&\sum_{j=1}^\infty
g_{1,2}^j(\tilde p)\,X_{2j-2,0}(\chi_p),\label{s3p1}\\
g_{3,4}(p_4,|\bp|)&=&\sum_{j=1}^\infty
g_{3,4}^j(\tilde p)\,X_{2j-1,1}(\chi_p),\\
g_{5,6}(p_4,|\bp|)&=&\sum_{j=1}^\infty
g_{5,6}^j(\tilde p)\,X_{2j,2}(\chi_p),\\
g_{7,8}(p_4,|\bp|)&=&\sum_{j=1}^\infty
g_{7,8}^j(\tilde p)\,X_{2j,1}(\chi_p),
\label{s3p2}
\end{eqnarray}
where the corresponding equations for the coefficients $g_n^j(\tilde p)$
can be readily obtained by inserting eqs. (\ref{s0p1})-(\ref{s3p2})
into eq. (\ref{par2s0}) and by using the orthonormalization relation
for $X_{jl}(\chi_p)$, eq.
(\ref{orto}).
In the $^1S_0$ channel
 the resulting equations are
\begin{eqnarray}
g^j_{1,2}(\tilde p)&=&-g^2\ b_{1,2}\int\limits_{0}^\infty
\frac{d\tilde k\, {\tilde k}^3}{8\pi^2(2j-1)}\,V_{2j-2}(\tilde p,\tilde k)
\sum\limits_{n=1}^4\sum\limits_{m=1}^\infty
A^{1,2\ n}_{jm}(\tilde k)\,g_n^m(\tilde k),
\label{set1}\\
g^j_3(\tilde p)&=&-g^2b_3 \int\limits_{0}^\infty\frac{d\tilde k\,
{\tilde k}^3}{8\pi^2\, 2j}\,V_{2j-1}(\tilde p,\tilde k)
\sum\limits_{n=1}^4\sum\limits_{m=1}^\infty
A^{3\ n}_{jm}(\tilde k)\,g_n^m(\tilde k),
\\
g^j_4(\tilde p)&=&-g^2 b_4\int\limits_{0}^\infty\frac{d\tilde k\,
{\tilde k}^3}{8\pi^2(2j+1)}\,V_{2j}(\tilde p,\tilde k)
\sum\limits_{n=1}^4\sum\limits_{m=1}^\infty
A^{4\ n}_{jm}(\tilde k)\,g_n^m(\tilde k).\label{set4}
\end{eqnarray}
For the $^3S_1$--$^3D_1$ channel we get
\begin{eqnarray}
g^j_{1,2}(\tilde p)&=&-g^2 b_{1,2}\int\limits_{0}^\infty
\frac{d\tilde k\, {\tilde k}^3}{8\pi^2(2j-1)}\,V_{2j-2}(\tilde p,\tilde k)
\sum\limits_{n=1}^8\sum\limits_{m=1}^\infty
B^{1,2\ n}_{jm}(\tilde k)\,g_n^m(\tilde k),\label{set11}\\
g^j_{3,4}(\tilde p)&=&-g^2 b_{3,4}\int\limits_{0}^\infty
\frac{d\tilde k\, {\tilde k}^3}{8\pi^2\, 2j}\,V_{2j-1}(\tilde p,\tilde k)
\sum\limits_{n=1}^8\sum\limits_{m=1}^\infty
B^{3,4\ n}_{jm}(\tilde k)\,g_n^m(\tilde k),\\
g^j_{5-8}(\tilde p)&=&-g^2 b_{5-8}\int\limits_{0}^\infty
\frac{d\tilde k\, {\tilde k}^3}{8\pi^2(2j+1)}\,V_{2j}(\tilde p,\tilde k)
\sum\limits_{n=1}^8\sum\limits_{m=1}^\infty
B^{5-8\  n}_{jm}(\tilde k)\,g_n^m(\tilde k),
\label{set14}
\end{eqnarray}
where the explicit expressions for
$A^{i n}_{jm}(\tilde k)$ and $B^{i n}_{jm}(\tilde k)$
are presented  in Tables \ref{tab7}, \ref{tab8} and \ref{tab9}
(for  details see    Appendix \ref{app:kernel}). This is our
main analytical result.
As seen from Tables \ref{tab5} and \ref{tab6}, the partial amplitudes
$g_3$ in the $^1S_0$ channel and $g_2$, $g_6$ and $g_8$
in the $^3S_1-^3D_1$ channel are identically zero
for the vector meson exchange, which
is a direct consequence of our (approximate) choice of the
propagator of vector particles.

An implementation of the vertex form factors (\ref{ff}) into the calculation
 is straightforward. Namely, by observing that in  Euclidean space
\begin{eqnarray}
\frac {1}{(p-k)^2+\mu^2}\, F[(p-k)^2]^2=2\pi^2 \sum_{nlm}
\frac {1}{n+1}\, \tilde V_n(\tilde p,\tilde k)\,Z_{nlm}(\chi_p, \theta_p,\phi_p)
\, Z_{nlm}^{*}(\chi_k, \theta_k,\phi_k),\label{e1}
\end{eqnarray}
with
\begin{eqnarray}
\tilde V_n(\tilde p,\tilde k)&=&  4\left[ \frac {(\Lambda_+^{\mu}
-\Lambda_-^{\mu})^n}{(\Lambda_+^{\mu} +\Lambda_-^{\mu})^{n+2}}
-\frac {(\Lambda_+^{\Lambda}
-\Lambda_-^{\Lambda})^n}{(\Lambda_+^{\Lambda}
 +\Lambda_-^{\Lambda})^{n+2}} \right]
 \nonumber \\[3mm]
 &-&4(n+1)\,
 \frac{\Lambda^2-\mu^2}{\Lambda_+^{\Lambda}\Lambda_-^{\Lambda}}\,
 \frac {(\Lambda_+^{\Lambda} -\Lambda_-^{\Lambda})^n}{(\Lambda_+^{\Lambda}
 +\Lambda_-^{\Lambda})^{n+2}},
 \label{e2}\\[3mm]
\Lambda_\pm^{\mu}&=&\sqrt{(\tilde p\pm \tilde k)^2+\mu^2},\quad
\Lambda_\pm^{\Lambda}=\sqrt{(\tilde p\pm \tilde k)^2+\Lambda^2},
\nonumber
\end{eqnarray}
one finds that the only modification
of eqs. (\ref{set1})- (\ref{set14}) consists in replacing
the quantities $V_n(\tilde p,\tilde k)$ of (\ref{vn}), by the "dressed" quantities
$\tilde V_n(\tilde p,\tilde k)$ of (\ref{e2}).
Hence, one can conclude that the system of equations
(\ref{set1})- (\ref{set14}) together
with the proposed new basis  (\ref{nharms})
and (\ref{nharmd}) present a direct generalization of the
hyperspherical harmonics method for the spinor-spinor BS equation.
Since the new basis used here is less intuitive
than the more traditional  $\rho$-spin basis, one
would like to relate the results to the $\rho$-spin formalism where
the computed matrix elements have a clear physical meaning
and allow for direct comparisons with nonrelativistic calculations.
This can be done by first solving the BS equation for the
partial components $g_n$ and then by
 use of eqs. (\ref{1s0}) and (\ref{3s1})  finding
 the desired partial components $g_\alpha$ of the $\rho$-spin formalism.

\section{Numerical solutions}
   Equations (\ref{set1})-(\ref{set14}) represent the desired system
   of the BS one-dimensional integral equations within the hyperspherical
   harmonics formalism to be solved numerically.
   Before choosing a specific computational
   algorithm one has to analyze at least  two issues,
i) existence and uniqueness of the solution and ii) if
   such a solution exists, one has to analyze
    the convergence and stability of the  (approximate) solution to the exact one.
   Obviously, both issue are tightly connected
   to the properties of the interaction kernels.
    The main   requirement for the existence of solutions of the Fredholm type
     equations is the finiteness of the kernel.
    This can always be  fulfilled by  considering the
    cut-off form factors (\ref{ff}) in the  interaction Lagrangians.
    To find the numerical solution,  one transforms the
    continuous space of arguments and solutions into a discrete one,
    forms the skeletons of the approximate kernel and of the solution, chooses
    a numerical method to solve the resulting (finite) system of linear equations
     and investigates, within such a scheme,
     the convergence of the skeletons to their exact originals.
      In such a way one estimates the effectiveness and correctness of the
      algorithm of the chosen procedure.
      However,  inclusion
      of cut-off form factors is not always  physically justified. Moreover,
      in some specific cases  the interaction kernel can increase unbound
      and therefore, does not automatically guarantee  the existence of a solution.
      Within nonrelativistic quantum mechanics it is known that
      the bound state of the Schr\"odinger equation with the interaction potential of
      the form $U(r)=-\alpha/r^2$ ($r$ is the radius vector in  co-ordinate space)
      can disappear~\footnote{This effect is known as collapse, i.e.
      when the particle classically "falls" into the center}
      at large enough coupling constant $\alpha$~\protect\cite{LL}.
     A similar situation occurs also in relativistic quantum mechanics
     within the Light Front Dynamics~\cite{karmanov,karm2,gl},
     where  for potentials corresponding to
     exchanges of scalar mesons, a critical value $g_\mathrm{cr}$
    of the coupling constant exists,
     above which the bound state disappears. Recall that  this refers to the case
      without cut-off form factors. Inclusion   of
     form factors, e.g. eq.~(\ref{ff}),  essentially aims to improve
     convergence of the system of equations and to assure the existence
     of a solution for any type of the  (attractive) exchange meson.

\subsection{Stability  of solutions }
In this section we investigate the  existence and stability of the
solution of the  BS equation without cut-off form factors
for two spinor particles.
For definiteness, we  consider  the system (\ref{par1s0}) in the
$^1S_0$ channel for the  scalar exchange.

Since we are interested in clarifying the existence of a critical value
of the coupling constant $g_\mathrm{cr}$, the system (\ref{par1s0}) is analyzed
at asymptotically large values of $\tilde p$. In this case it is sufficient to
investigate the properties of the
system (\ref{par1s0}) in the two-dimensional space ($\tilde p,\chi_p$),
  without further decomposition (\ref{s0p1})-(\ref{s0p1})
 into Gegenbauer polynomials.
The equation
for the component $g_1$ is then
\begin{eqnarray}
g_1({\tilde p},\chi_p)=\frac {g^2}{(2\pi)^4}\int
\frac {d{\tilde k}\, {\tilde k}^3 d\chi_k \sin^2 \chi_k\, d\Omega_k
}{({\tilde k}^2+m^2-M^2/4)^2+M^2 {\tilde k}^2\cos \chi_k}\, \frac {1}{(p-k)^2+\mu^2}
\label{g1only}\\
\nonumber\times
[(M^2/4+m^2+{\tilde k}^2)\,g_1({\tilde k},\chi_k)+Mm\, g_2({\tilde k},\chi_k)-M{\tilde k}
\sin \chi_k\, g_3({\tilde k},\chi_k)].
\end{eqnarray}
For further convenience
we introduce a new integration variable $\gamma$ by ${\tilde k}=\gamma {\tilde p}$.
Then at $\tilde p\to\infty$  in eq. (\ref{g1only})  only terms  proportional
to $\tilde k^2$ survive and  eq. (\ref{g1only}) becomes
\begin{eqnarray}
g_1({\tilde p}, \chi_p)= \frac {g^2}{(2\pi)^3} \int
\limits_0^{\infty} d\gamma \int \limits_0^{\pi} \sin^2
\chi_k d \chi_k \int \limits_{-1}^1 d\cos\theta_k
V(\gamma,\cos\theta_k,\chi_p,\chi_k)\,g_1(\gamma {\tilde p},
\chi_k), \label{eqk2}
\end{eqnarray}
where the interaction kernel
\begin{eqnarray}
V(\gamma,\cos\theta_k,\chi_p,\chi_k)=\frac
{\gamma}{1+\gamma^2-2\gamma \cos \chi_p \cos \chi_k-2\gamma
\cos\theta_k \sin \chi_p \sin \chi_k}. \label{eqk3}
\end{eqnarray}
is positively defined and  the relation
$V(\gamma,\cos\theta_k,\chi_p,\chi_k)=V(1/\gamma,\cos\theta_k,\chi_p,\chi_k)$ holds.
From eqs. (\ref{eqk2}) and (\ref{eqk3}) one infers that the integral converges
if $g_1(\tilde p,\chi_p)$ asymptotically vanishes
as $1/\tilde p$ or faster, i.e.  if $g_1$ can be written in the form
\begin{eqnarray}
g_1({\tilde p},\chi_p)= \frac{h(\tilde p,\chi_p)}{\tilde p^{1+\beta}\sin\chi_p},
\label{asf}
\end{eqnarray}
where $\sin\chi_p$ has been introduced for  convenience.
Now, by  splitting  the integration over
$\gamma$ in (\ref{eqk2}) into two sub-ranges as
\begin{eqnarray}
\int \limits_0^{\infty} \dots d\gamma =\int \limits_0^{1} \dots
d\gamma +\int \limits_1^{\infty} \dots d\gamma,
\label{spl}
\end{eqnarray}
and changing the variable in the second integral  $\gamma\to
1/\gamma$ and carrying out integration over $\cos\theta_k$ analytically,
 we obtain for (\ref{eqk2})
\begin{eqnarray}
  \int \limits_0^1 \frac{d\gamma}{\gamma}\,
\int \limits_0^{\pi}  d\chi_k\ \cosh(\beta\ln\gamma)
\ln \frac{1 -a \cos(\chi_p+\chi_k)}{1-a\cos(\chi_p-\chi_k)}
\, h(\tilde k,\chi_k)
=\frac{(2\pi)^3}{g^2}\,h(\tilde p,\chi_p),
\label{eq3.4}
\end{eqnarray}
where $a=\displaystyle\frac{2\gamma}{1+\gamma^2}$.
Eq. (\ref{eq3.4}) can be considered as an equation of the Sturm-Liouville-like
problem of finding the eigenvalues $\displaystyle\frac{(2\pi)^3}{g^2}$
and the eigenvectors $h(\tilde p,\chi_p)$ of the corresponding integral operator.
It is immediately seen that the eigenvalues depend on the asymptotic
behavior of $g_1$ and that the most harmful situation occurs
at $\beta=0$ and $h(\tilde p,\chi_p)=h(\chi_p)$.
Namely, for $\beta <0$ the integral is divergent and the equation becomes meaningless,
for $\beta >0$ the integral converges rather fast ensuring the existence of bound states.
Thereby, the critical value of the
 coupling constant $g_\mathrm{cr}$  can occur only
 at $\beta=0$. Note, that from eq. (\ref{eq3.4}) one finds that the function
 $h(\tilde p,\chi_p)$ is odd with respect  to  the variable $ \chi_p $,
$ h(\tilde p,-\chi_p)=-h(\tilde p,\chi_p)$. This implies that
at $\beta=0$ and $h(\tilde p,\chi_p)=h(\chi_p)$ the quantity $h(\chi_p)$ can be
developed into an odd Fourier series

\begin{eqnarray}
  h(\chi_p)=\sum\limits_{n=1}^\infty c_n
  \sin (n\chi_p).
  \label{solh}
\end{eqnarray}
It is straightforward to check  that  each term of
the series (\ref{solh}) represents a solution of the eq. (\ref{eq3.4}) with $\beta=0$.
Moreover,   since
\begin{eqnarray}&&
\label{nalait}
  \int \limits_0^1 \frac{d\gamma}{\gamma}\,
\int \limits_0^{\pi}  d\chi_k\
\ln \frac{1 -a \cos(\chi_p+\chi_k)}{1-a\cos(\chi_p-\chi_k)}
\, \sin(n\chi_k)=\int\limits_0^1 \frac{d\gamma}{\gamma}\int\limits_0^{2\pi} d\xi
\frac{a}{n}\frac{\sin\xi}{1-a\cos\xi}
\cos(n(\xi-\chi_p))\nonumber\\&&
=\frac{a}{n}\int\limits_0^1 \frac{d\gamma}{\gamma}\,
\Re\left\{{\rm e}^{-in\chi_p}   \int\limits_0^{2\pi} d\xi \frac{\sin\xi}{1-a\cos\xi}
{\rm e}^{in\xi}\right\}
=\frac{2\pi}{n^2}\sin(n\chi_p),
\label{eq3.42}
\end{eqnarray}
 the critical value  $g^2_\mathrm{cr}$ depends on $n$,
being $g_\mathrm{cr}=2\pi n$. Obviously,
the lowest   $g_\mathrm{cr}$ occurs at $n=1$, i.e. $g_\mathrm{cr}=2\pi$.

 Similarly one finds the critical value of the coupling constant from the
asymptotic
 equation for  $g_2$. The result is that the lowest critical value
 is determined by the solution of the form
 $\tilde h(\chi_p)=\sin \chi_p +\epsilon \sin 3\chi_p$
(with $\epsilon=1.100925$), which
 provides
$g_\mathrm{cr}^2 \simeq 71.71862$. The remaining components $g_3$ and $g_4$
are negligibly
small and can be neglected in the present analysis.
Hence, we find that in the $^1S_0$ channel the critical value
 $g_\mathrm{cr}^2\sim 4\pi^2$.  This
 is confirmed, with good accuracy, by concrete numerical
 calculations  where  no bound states for
$g > g_\mathrm{cr}$ are found at
 ${\tilde p}\rightarrow \infty$. This is also in full accordance with
the nonrelativistic potential $V(r)\simeq -\alpha/r^2$.

 The critical value
for the coupling constant for the   $^3S_1-^3D_1$ channel
above which the bound state disappears is
 $g^2_\mathrm{cr}\sim 78$.

\subsection{Numerical methods}
In concrete numerical calculations we form the skeletons of approximate
solutions and kernels by using the Gaussian   method of computing  integrals
and by restricting the infinite sum over $m$ in   eqs. (\ref{set1})-(\ref{set4})
and (\ref{set11})-(\ref{set14})   by
a finite value $M_\mathrm{max}$.
The Gaussian quadrature formula assures a rather good convergence of the
numerical procedure  and provides the
sought solution in the Gaussian nodes which are spread rather uniformly in the
 interval $0\le \tilde p < \infty$. In order to have the solution in  detail
at moderate values of $\tilde p$, which is the interval of the actual physical interests,
one usually redistributes the Gaussian mesh making
the nodes  more dense at low values of $\tilde p$.
To this end one  applies an appropriate mapping of the Gaussian mesh
by  changing  of variables as~\cite{fb-tjon}
\begin{eqnarray}
\tilde p=\tilde p(x)=c_0\,\frac{1+x}{1-x}\label{varch}
\end{eqnarray}
with  $c_0$ as a free parameter and $-1\le x\le 1$.
Then the corresponding    set of linear equations reads as
\begin{eqnarray}
X=g^2\, AX,\label{syst}
\end{eqnarray}
where  the vector
\begin{eqnarray}
X^T=\left ([\{g_1^m({\tilde k}_i)\}_{i=1}^{N_G}]_{m=1}^{M_\mathrm{max}},
[\{g_2^m({\tilde k}_i)\}_{i=1}^{N_G}]_{m=1}^{M_\mathrm{max}},
\ldots,
[\{g_{n}^m({\tilde k}_i)\}_{i=1}^{N_G}]_{m=1}^{M_{max}}\right )
\label{vector}
\end{eqnarray}
represents the sought solution in the form of a group of sets of
partial wave components $g_n^m, n=1,\ldots,n_\mathrm{max};
m=1,\dots,M_\mathrm{max}$ specified on the integration mesh of the order $N_G$.
As before, in eq. (\ref{vector})  $n_\mathrm{max}=4$
and  $n_\mathrm{max}=8$ for the $^1S_0$ and $^3S_1-^3D_1$ channels,
respectively. The matrix $A$
is determined by   the corresponding  partial kernels,
the Gaussian weights
and the  Jacobian of the transformation (\ref{varch}) and is of the
 $N\times N$ dimension, where $N=n_\mathrm{max}\times M_\mathrm{max}\times N_G$.
Since the system of equations (\ref{syst}) is homogeneous,
the  eigenvalues $g^2$ (at given  mass of the bound state $M$)
are obtained from the condition
$\det(g^2 A-1)=0$. Then
the partial components $g_n^m$ are
found by solving  numerically
the system (\ref{syst}) with these  eigenvalues $g^2$.

We use a combined method of finding the solution $X$.
 First, the Gauss-Jordan elimination and pivoting method
  involving the choice of the leading element
 \cite{forsyth} is applied.
  Then the obtained solution is used as a trial input into an
   iteration procedure to find (after $ 5-10 $ iterations)
  more refined results. Within such an algorithm
 we investigated the convergence of the approximate solution
 by increasing the dimension of the matrix $A$ and found that   the
 method is stable and robust up to values $N\sim 4000$. This is quite enough
 to obtain solutions with practically any desired accuracy.

Another way to solve the system (\ref{syst}) consists in directly using
the iteration method with known nonrelativistic wave functions as
trial inputs. This method has been  inspired by
the success  of the One Iteration
Approximation scheme developed in \cite{dp,our_chex,our_brkp},
which provides a quick and accurate solution.
In the present paper both of these methods are widely explored. In
practice, however, instead of finding the eigenvalues $g^2(M)$  at given
 $M$, one  usually considers the inverse problem when
 the coupling constant $g^2$ is kept   fixed
and  the mass $M(g^2)$ is assumed  as a function of $g^2$.

As an illustration of the stability
of the numerical procedure, in Table~\ref{tab10}
we present results for the   masses of the bound state
$M(g^2)$  depending on the Gaussian mesh $N_G$
 and $M_\mathrm{max}$. Calculations have been performed for the
$^1S_0$ state, eqs. (\ref{set1})-(\ref{set4}), with a scalar
meson exchange of  mass $\mu$ for two
values   $\mu=0.15\ \mathrm{GeV/c}^2$  and $\mu=0.5\ \mathrm{GeV/c}^2$;
the constituent particles (nucleons) have been taken
with equal masses  $m= 1.0\ \mathrm{GeV/c}^2$ for simplicity.
 Results presented in Table~\ref{tab10} clearly demonstrate
 that the approximate solution
 converges rather rapidly, and already
at $M_\mathrm{max}\sim 4-5$ and $N_G=64$ the method provides  a good solution of the
system (\ref{set1})-(\ref{set4}). Obviously, the coupling constant
$g^2$ must be taken small enough, i.e. $g^2 < g^2_\mathrm{cr}\sim 40$ to ensure
the existence of the solution.
The free parameter $c_0$ in (\ref{varch}) does not affect the
convergence and for definiteness it has been set $c_0=1$.
As expected, an implementation  of the cut off form
factors (\ref{ff}), (\ref{e1}) and (\ref{e2})
essentially improves the convergence of the approximate solution.

\begin{figure}[ht]             
~\centering
\epsfxsize 4.5in
\epsfbox{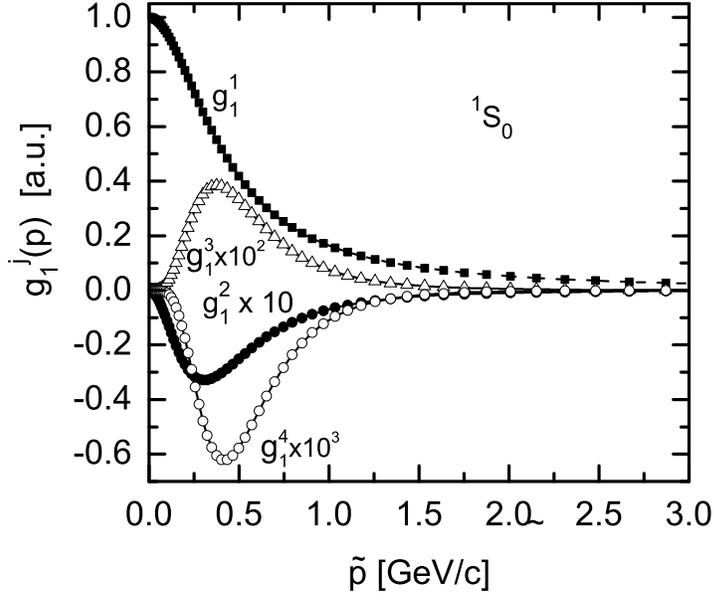}
~\vskip -6mm
\caption{Functions $g_1^j, j=1,\ldots,4$, eq.  (\ref{s0p1}),
  at the gaussian  mesh points $N_G=96$. Closed squares correspond to
$g_1^1$,  closed circles - $g_1^2$ multiplied by 10,
triangles -$g_1^3$ multiplied by 100, open circles -$g_1^4$ multiplied by 1000,
the solid lines correspond to the  fitted functions  $g_1^j $
 by formula (\ref{fit}).
The overall   normalization constant is arbitrary.}
\label{pic1}
\end{figure}

 To display the behavior of the  components
 of the vertex function,  in Fig.~\ref{pic1}
  we present the resulting coefficients  for the $^1S_0$ channel
  $g_1^j(\tilde p)$, $j=1 \ldots 4$, eq.  (\ref{s0p1}),
  at $M=1.937\ \mathrm{GeV/c}^2$ for $N_G=96$, $M_\mathrm{max}=4$ and $g^2=15$.
 It can be seen that at large $\tilde p$  these functions decrease
as inverse powers of $\tilde p$ which permits to cast the approximate
solution into the form
\begin{eqnarray}
\label{fit} g_1^j(\tilde p)\simeq
 \left[\frac{\tilde p^2}{\tilde p^2+b_j^2}\right]^{j-1}
\sum\limits_{l=1}^4\frac{a_{jl}\,\tilde p^{2l-2}}{(\tilde p^2+b_j^2)^l},
\end{eqnarray}
\noindent
where the parameters $b_j$ and $a_j^l$ can be found
from an $\chi^2$ analysis  of  the  adjustment of  eq.~(\ref{fit})
to the approximate solution (see Table \ref{tab11}).
The solid lines in Fig.~\ref{pic1} reflect the result
of the fit of the   numerical solution by
eq.  (\ref{fit}). The  accuracy of the
results implies  that in such a way one
can find solutions of the corresponding system of equations
as continuous functions of $\tilde p$ which are  extremely
useful in practical applications. Similar  analysis of other
coefficient functions $g_{2,3,4}^j(\tilde p)$
shows that an excellent fit of the numerical
solutions can be achieved  for all the partial components $g_\alpha(\tilde p)$.
From this encouraging result we argue
that this method using hyperspherical harmonics can be considered as a  reliable
tool to solve the Bethe-Salpeter equation numerically, even with different
parameterizations  of the solution than the simple form (\ref{fit}).
Unfortunately, for small meson masses $\mu\sim 0$ the method becomes less effective,
having a poor or even failing  convergence feature
and  hence becomes less adequate, requiring  a separate analysis
(see also Ref.~\cite{fb-tjon}).

\begin{figure}[ht]             
~\centering
\epsfxsize 4.5in
\epsfbox{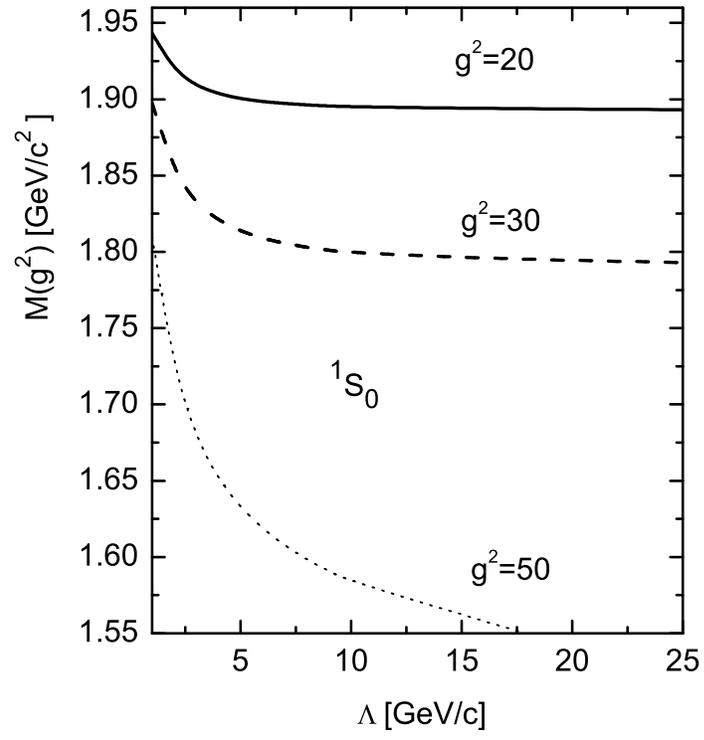}
~\vskip -6mm
\caption{Mass of the bound state in the $^1S_0$ channel as a
functions of the cutoff parameter $\Lambda$ for different values
of the coupling constant $g$ at
$m=1\ \mathrm{GeV/c}^2$, $\mu=0.15\ \mathrm{GeV/c}^2$}
\label{pic2}
\end{figure}

\begin{figure}[ht]                   
~\centering
\epsfxsize 4.5in
\epsfbox{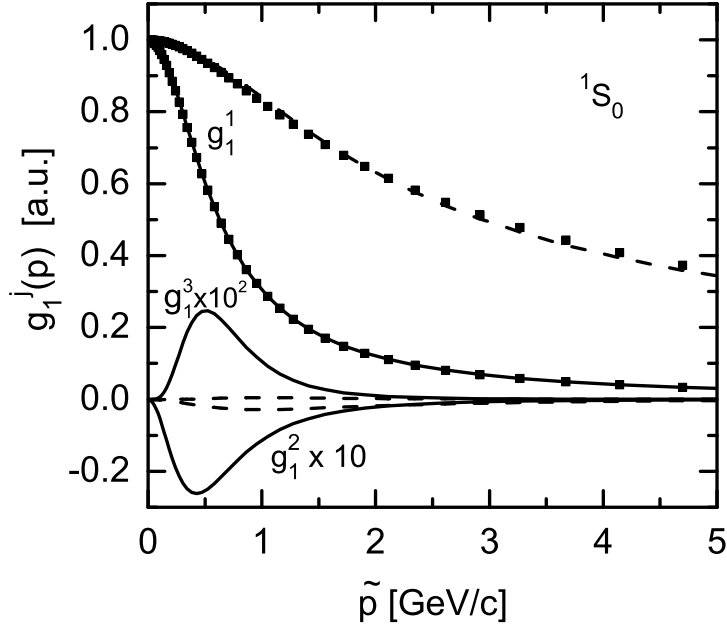}
~\vskip -6mm
\caption{Hyperspherical components $g_1^j, j=1,2,3$, for
 $\Lambda=5\ \mathrm{GeV/c}$  and  coupling constant
$g^2=30$ (solid line) and $g^2=48$ (dashed line). For the
component $g_1^1$ the lines reproduce the results of interpolation
by eq. (\ref{fit1}).}
\label{pic3}
\end{figure}

\begin{figure}[ht]                  
~\centering
\epsfxsize 4.5in
\epsfbox{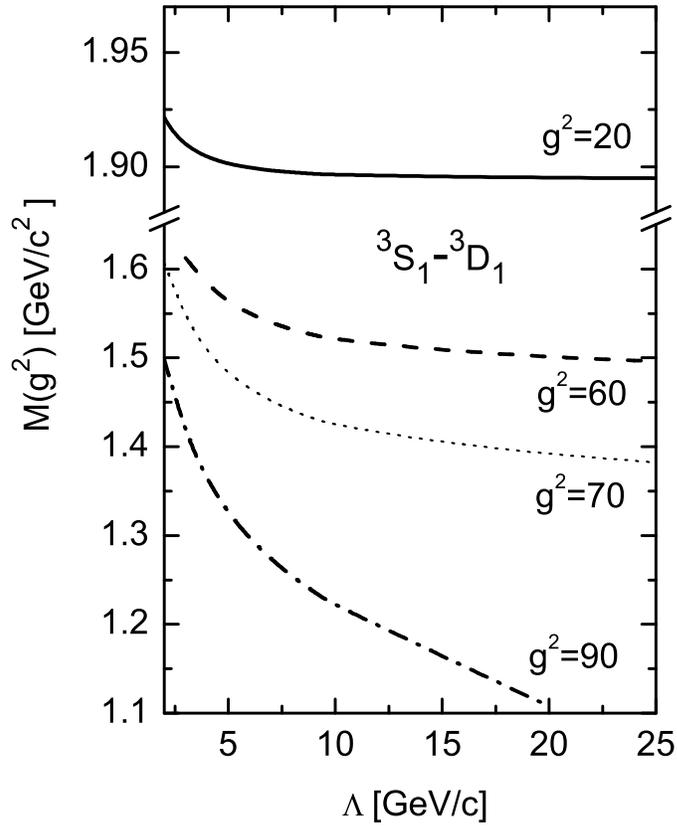} ~\vskip -6mm
\caption{Masses of bound state in the $^3S_1$-$^3D_1$ channel as
functions of the cutoff parameter $\Lambda$ for different values
of the coupling constant $g$.
 Critical value $g^2_\mathrm{cr}\sim 78$. }\label{pic4}
\end{figure}

Also, as mentioned above, at large
values of the coupling constants
$g^2$, close to or even larger than the critical value $g^2_\mathrm{cr}$,
the numerical solutions become strongly dependent
on $M_\mathrm{max}$,  $N_G$ and $c_0$ which is a clear signal that
(in absence of the cut-off form factors (\ref{ff}))
 the solution becomes unstable at
$g^2\sim g^2_\mathrm{cr}$ and disappears at $g^2 > g^2_\mathrm{cr}$.
Such a situation is illustrated in  Fig. \ref{pic2}
where we present the dependence of the mass $M$  on the values of the
cut-off parameter $\Lambda$ at different  coupling constants
$g^2$, below and above the critical value $g^2_\mathrm{cr}\sim 40$.
It is seen that if the solution exists and is stable ($g^2 <g^2_\mathrm{cr}$)
then the mass $M$ is almost independent on the cut-off form factor
and tends to a constant value at large $\Lambda$.
Contrarily, the dependence of the mass $M$ on
  $\Lambda$ at  $g^2 > g^2_\mathrm{cr}$
evidently indicates that in this case the solution becomes  unstable
and at large  values of $\Lambda$, $\Lambda\to\infty$, it can disappear at all.
Such a behavior of the  solution at $g^2\sim g^2_\mathrm{cr}$
 exactly reproduces the peculiarities
 of the well-known
 collapse phenomenon for  potentials like $-\alpha/r^2$ in
 nonrelativistic quantum mechanics (see, also discussions in Ref.~\cite{karm2}).
In Fig. \ref{pic3} the numerical solutions for
the partial components $g_1^j, j=1,2,3$
are given for two values of the coupling constant,
below  (solid lines) and above the critical $g_\mathrm{cr}$ (dashed lines).
In order to ensure the existence of the solution, calculations have been performed at
a finite, however large,  value of the cut-off parameter $\Lambda=500\ \mathrm{GeV/c}$.
It is seen that in the case of $g^2 > g_\mathrm{cr}^2$ the asymptotic
decrease of the wave functions is rather weak which implies the
  instability of the solution. In Fig. \ref{pic3} the continuous
lines reflect the result of a fit   by eq. (\ref{fit}) which   for
 the asymptotic region of the
component $g_1^1$,   can be
written in a simple form
\begin{eqnarray}
\label{fit1}
g_1^1(\tilde p)\to \frac{a}{(\tilde p^2+b^2)^c},
\end{eqnarray}
where $c=0.77$ ($g<g_\mathrm{cr}$) and  $c=0.4$ ($g>g_\mathrm{cr}$).
A comparison with eqs. (\ref{asf}) and (\ref{solh})
  shows that for  coupling constants below the critical value,
  $g<g_\mathrm{cr}$,  one has $\beta >0$, while above $g<g_\mathrm{cr}$,
    $\beta < 0$, i.e., the  integral (\ref{eqk2}) without   form factors (\ref{ff})
 diverges.
 A similar behavior of the numerical solution
occurs  in the $^3S_1$-$^3D_1$ channel as well (see, Fig.~\ref{pic4}).

\subsection{Scalar coupling}
Having established the main features of the numerical
procedure, we solve the BS equation within the hyperspherical harmonics method
for different kinds of the exchanged mesons, i.e. scalar, pseudoscalar and
vector mesons. We investigate the dependence of the bound state mass $M$
on the coupling constant $g$ and study the partial components
$g_\alpha$ as a function of the hypervariable $\tilde p$ at fixed values of
$M$ and the coupling constant $g$.
All calculations have been done with and without the
cut-off form factors (\ref{ff}) and,
obviously, for  the coupling constants below their critical values.
We compare our results with other calculations
performed for similar conditions, namely we widely  compare our analysis to
the ones  obtained by Light Front (LF)
dynamics~\cite{karmanov} and by the non relativistic (NR)
Schr\"odinger approach. Note that within these approaches
the  dynamical variables used differ slightly from the ones
 used in BS formalism, hence a direct comparison of the calculated quantities
is hampered. However, one can reconcile approaches by
choosing one variable within the BS formalism,
e.g. the modulus of the 3-dimensional relative momentum $\bf p$,
 and relate the corresponding quantities
 in LF or NR approaches through this variable for further
 comparisons.

\begin{figure}[ht]                   
\centering
\epsfxsize 4.5in
\epsfbox{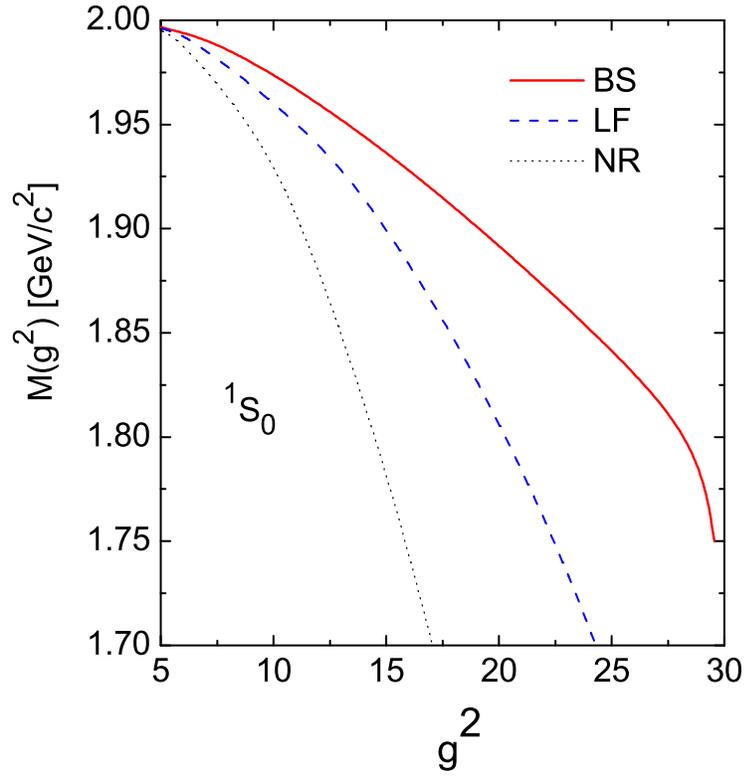}
~\vskip -6mm
\caption{Dependence of the mass of the bound state
$M$ in the $^1S_0$ channel for
different approaches. The solid line represents
results  within the BS  formalism,  dashed line corresponds to
Light Front (LF) dynamics
and the dotted line is the result of the non relativistic (NR)
Schroedinger approach.
}
\label{pic5}
\end{figure}

\begin{figure}[ht]             
~\centering \epsfxsize 4.5in
\epsfbox{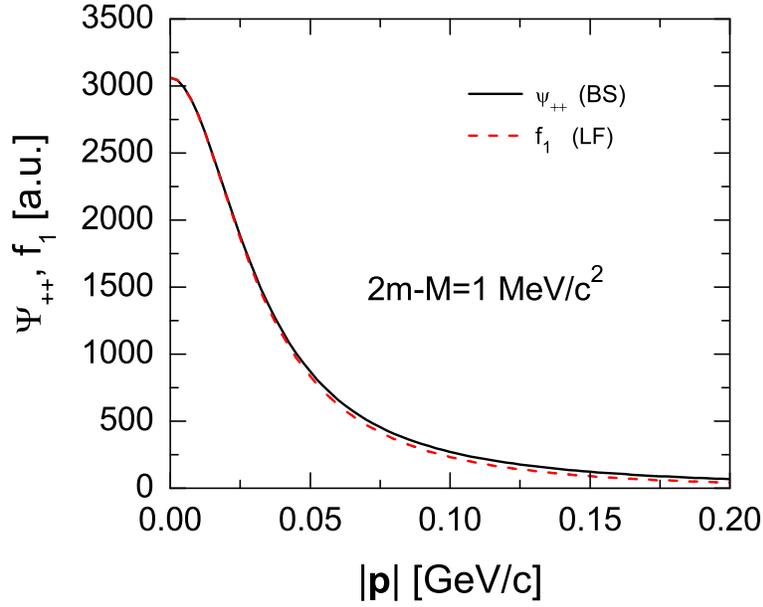}
~\vskip-6mm
\caption{Comparison of the BS partial "++"-component
eq. (\protect\ref{ppp}), solid line, with LF
partial component $f_1$ Ref.~\protect\cite{our-BS-LF}, dashed line,
at the binding energy $B \equiv 2m-M=1\ \mathrm{MeV/c}^2$.}
 \label{pic6}
\end{figure}

\begin{figure}[ht]                  
~\centering
\epsfxsize 4.5in
 \epsfbox{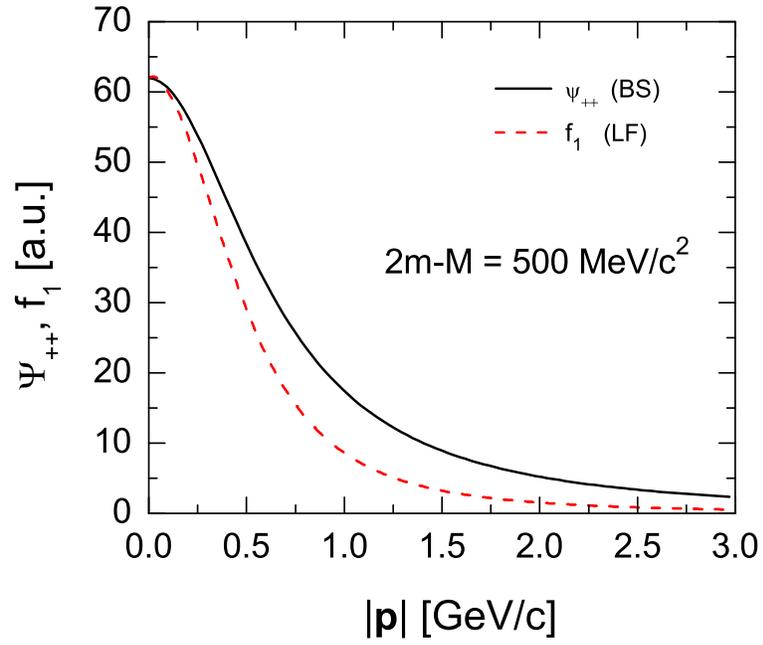}
 ~\vskip -6mm
 \caption{The same as in Fig. \ref{pic6} but for the binding
energy $B \equiv 2m-M=500\ \mathrm{MeV/c}^2$. }
\label{pic7}
\end{figure}

 Such relations have been found  and
 reported in detail in Refs.~\cite{our-BS-LF,karmanov_PR}. Here
 we note only that in determining  relations
 between the BS and LF amplitudes one finds that in the $^1S_0$ channel
 only two components ($^1S_0^{++}$ and $^3P_0^e$)
 of the BS amplitude correspond pairwise to the two
 LF  functions $f_1$ and $f_2$ (notation as in Ref. ~\cite{karmanov}).
The other BS components, when being projected
on to the LF surface, vanish. Similarly, in the $^3S_1-^3D_1$ channel
only five BS components survive on to LF surface~\cite{our-BS-LF,karmanov_PR}.
In Fig.~\ref{pic5}  the mass of the bound state as a function of the
coupling constant for scalar meson exchange is shown.
The solid line corresponds
to results within the  BS approach, while the dashed and dotted lines
show  calculations with LF~\cite{karmanov} and
NR (with  an Yukawa-type potential) approaches, respectively.
Only at low values of the binding energy
different approaches provide similar results. As the coupling
constant increases (increase of the binding energy)
 the difference becomes more and more significant.   Even within
 two relativistic frameworks, BS and LF formalisms, the difference
 increases with the binding energy increase. This is illustrated also
in Figs.~\ref{pic6} and~\ref{pic7} where the wave function
$\Psi_{++}$, eq. (\ref{ppp}), is compared to the corresponding
LF wave function $f_1$~\cite{karmanov} for two different binding energies.
It is seen that the two approaches provide
similar results for low,  and rather different results
for high values of the binding energy.
 Obviously, this is a direct consequence
 of the different treatment of the relativistic effects
 within the BS and LF formalisms. Since when
 projecting  the BS  amplitude on to the LF
 surface, some partial components (with negative $\rho $-spins)
 vanish, the observed large difference between BS and LF results
serves as a hint that in this case the role
 of the  components with negative $\rho $-spins increases.
This can be checked  by computing the  pseudo probabilities for different
components,
eq.~(\ref{pseudopr}), at several values of the
binding energy. To this end, we  apply the
 transformation eq.~(\ref{1s0}) (or eq.~(\ref{3s1})
for the $^3S_1-^3D_1$ state) to the set of functions $g_n$, preliminarily
  obtained numerically from  the series
(\ref{s0p1})-(\ref{s0p2}) (or (\ref{s3p1})-(\ref{s3p2})).
\noindent
The corresponding results for the $^1S_0$ channel
are collected  in Table \ref{tab13}, where  the pseudo
probabilities for different components have been
computed  assuming the
total normalization to be
\begin{eqnarray}
P_{++}+P_{--}+P_{e}+P_{o}=1.\label{normir}
\end{eqnarray}
As expected, the contribution of the $^1S_0^{--}$ component
is negligibly small for weakly
bound systems and significant
for large binding energies.
 Similar conclusions hold  for the
$^3S_1$--$^3D_1$ state as well.
As mentioned, the BS system of equations for the partial
components have been solved for a coupling constant $g$ below
the critical value. However, there is also a lower limit below
which the solution of the BS equation disappears, in a fully analogy
with the  nonrelativistic Schr\"odinger equation,
when,  for shallow potentials, bound states do not exist.
Another observation is the result that the condition $\det(g^2 A-1)=0$
provides an equation for $M(g^2)$ which in the interval $0 < M < 2m$
can have more than one solution. In such a case the lowest value
of $M$ corresponds to the ground state, while others refer to the
discrete excited state of the system. We investigated also
the behaviour of the partial vertex functions $g_1,\dots,g_4$
for the excited states and found that, likewise in the nonrelativistic case,
these functions possess zeros as functions of  $|\bp|$.
This is in a good agreement with our previous
results obtained for the
two-dimensional mesh \cite{nash_yaf1}.

\subsection{Pseudoscalar meson exchange}

As  follows from eq. (\ref{par1s0}),  the system of equations
for pseudoscalar meson exchange is quite
similar to the previously described scalar case (see also  Tables \ref{tab2}-\ref{tab4}).
 However, contrary to the scalar case, when the main components
 $g_1$ and $g_2$ in eqs. (\ref{set1}) and  (\ref{set11})
are determined by pure attractive kernels,
for the pseudoscalar exchange
one of these components is governed by a pure repulsive kernel
 (c.f. Tables \ref{tab5} and \ref{tab6}).
Thus, the resulting balance of forces forming the
bound state for pseudoscalar exchange is more sophisticated.
As a consequence, to ensure an attractive residual kernel
for  creation of  a given bound state $M$, the BS equation
requires larger values of the coupling constant $g^2$. Consequently,
this can lead to subtle  situations when the minimal value of
$g^2$ is close to or even above the critical value $g_\mathrm{cr}^2$,
which may cause  problems with the stability of the
solution. Namely such a situation we  encountered in our numerical
calculations for a two-nucleon like system
 with pseudoscalar exchanges for which the stable bound state
(without cutoff form factors)  does not exist at all.
Similar result has been obtained also within
the LF approach~\cite{karmanov}
and seems to be of a general nature, i.e.
there is no relativistic bound state in a deuteron-like system
with pure pseudoscalar exchange.

\begin{figure}[ht]                  
  \centering
\epsfxsize 4.5in
\epsfbox{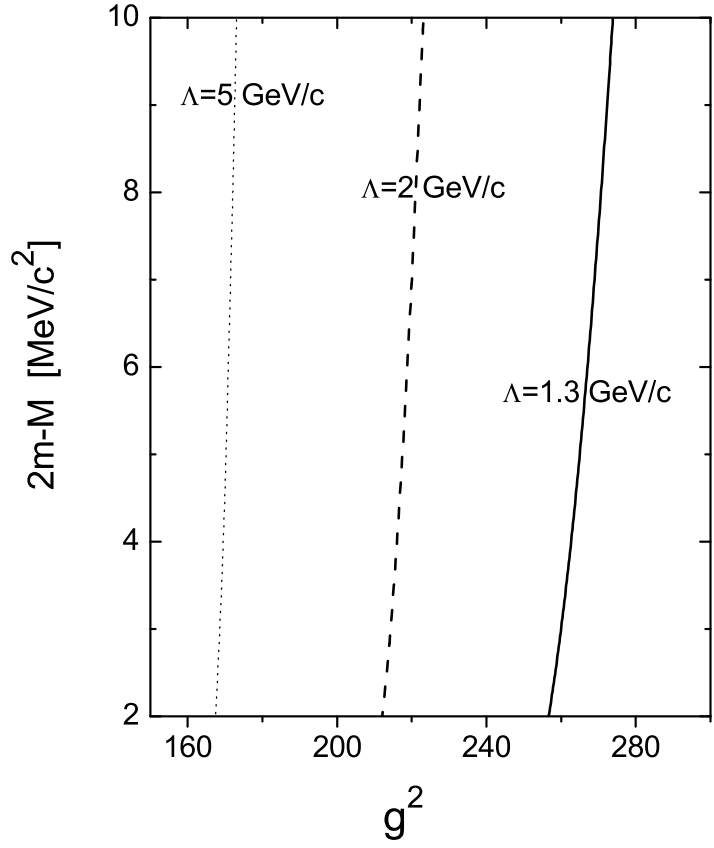}
\vskip -6mm
\caption{Sensitivity  of the binding energy
$B\equiv 2m-M$ to the coupling constant $g^2$ and  to
the cutoff parameter
$\Lambda$ for the $^1S_0$ state with pseudoscalar meson
exchange. }
\label{pic8}
\end{figure}

Another striking feature here is connected to the
behavior of the binding energy as a function of the coupling
constant. In Fig. \ref{pic8}, the  dependencies of the
binding energy   for the $^1S_0$ state are shown for
some values of the cutoff parameter $\Lambda$. It is clear that
such a dependence is very sharp -- binding energy increases very
rapidly with $g^2$, which is in agreement with the
results reported in Ref.~\cite{karmanov}.

\subsection{Vector meson exchange}
Since the contribution of vector meson exchange in
the nucleon-nucleon potential is repulsive
a deuteron-like bound state cannot be formed by pure vector
meson exchanges, therefore an investigation of the homogeneous
BS equation with such kernels is hampered. However, one can consider
a different two-fermion system such as the electron-positron pair, for
which the vector exchange potential does have a bound state.
Note that the general  form of the
BS equation, eq. (\ref{sphom}), in the particle-antiparticle channel
is maintained almost unchanged
(see, e.g. Refs. \cite{nakan} and \cite{itz}) so that
a relativistic description  of the positronium can be achieved by eq.
(\ref{sphom}) with  $\Gamma(1)=\gamma_{\mu}$,
$\tilde{\Gamma}(2)=\gamma_{\nu}$ and $V(p,k)=g_{\mu \nu}/(p-k)^2$.
Unfortunately, the knowledge of the
positronium, as a two-fermion bound system, is rather scarce~\cite{positr},
 even the absolute value of the binding energy is not definitely
established.

Nowadays only the transition energy between different
positronium levels (e.g., between para- and ortho-positronium) is an
object of experimental and theoretical investigation~\cite{positr}.
Compared to the electron mass this quantity is very small~\cite{LL}, being
of the order $\Delta B\sim m_e\alpha_\mathrm{em}^6/4 \sim
10^{-13}m_e$
(where  $\alpha_\mathrm{em}$ is the fine structure constant,
$m_e$ is the electron mass and $B\sim m_e\alpha_\mathrm{em}^2/4$
is the positronium binding energy predicted by the nonrelativistic
 Schr\"odinger equation). Consequently,
the procedure of solving the BS equation numerically for
the bound state $M\sim 2m_e-B$ and calculations of the
transition energy $\Delta B$ and comparison with
experimental data require extremely high
accuracy of calculations and large computational resources.
Moreover, as  mentioned before, in case of vanishing masses of
the exchange particle, the convergence of the method
is rather ill defined and the implementation of cut-off form factors becomes
a necessity in numerical calculations.
Therefore, in analyzing
effects of relativistic corrections  computed within different
schemes (e.g., within perturbative quantum electro-dynamics (QED),
LF dynamics and BS formalism) one usually solves the corresponding
equation with cut-off form factors and
at an effective coupling constant $g$ much larger than the
fine structure constant $\alpha_\mathrm{em}$ (see Ref. \cite{karmanov})
and compares the obtained results with those known analytically
 from the nonrelativistic Schr\"odinger equation.
For this reason, a commonly accepted value of $g$ is
$g^2=3.77$, which corresponds to
$\alpha=\frac{g^2}{4\pi}=0.3$ and to a nonrelativistic binding
energy $B=m_e\alpha^2/4=2.25 \cdot 10^{-2} m_e$~\cite{karmanov}.
In our calculations we  also adopted
such a coupling constant $g^2=3.77$
for which the para- and ortho- positronium
binding energy has been calculated. The   results are
presented  in  Table \ref{tab12}, where
the cut-off parameter  $\Lambda$ and binding energies  are given  in
units  of the electron mass $m_e$. The obtained results
demonstrate that at large enough values of
$\Lambda$, the solution of the BS equation is quite stable and almost independent
on $\Lambda$.
It is also seen that the BS equation provides
a  binding energy
almost twice  smaller than the nonrelativistic one which implies
that  the relativistic corrections
are of repulsive nature. Analogous result has been obtained within the LF dynamics
for the $0^-$ positronium~\cite{karm2}. However, calculations within the perturbative QED
show that the first order relativistic corrections are attractive.
This contradictory result can serve as  an indication that the adopted
interaction kernel is not accurate enough for a refined
description of the positronium within the  ladder approximation.
Other channels (e.g, the electron-positron annihilation~\cite{itz})
 and/or terms beyond the ladder approximation
should be included into the analysis.

\section {Conclusion}

We generalize a method based on hyperspherical harmonics
to solve the homogeneous spinor-spinor  Bethe-Salpeter equation
in  Euclidean space. To do so, we introduce a new basis of spin-angular harmonics,
suitable to expand the Bethe-Salpeter vertex into four-dimensional
hyperspherical harmonics. We obtain an explicit form of the corresponding system
of one-dimensional integral equations for the partial components
and formulate a proper  numerical algorithm to solve this system
of equations. The  BS vertex functions are  studied in detail for
  the $^1S_0$ and $^3S_1-^3D_1$ bound states with  scalar, pseudoscalar
  and vector meson exchanges.
Our results are in a good agreement  with calculations
within the  non relativistic and Light Front Dynamics approaches.

Within the novel method the effectiveness
of the numerical procedure is analyzed for the scalar, pseudoscalar
and vector meson exchanges and conditions for stability
of the solution  are established. It is demonstrated that above some critical
values of the coupling constant the solution of the BS equation
does not exist unless the cut-off form factors are considered.

 An advantage of the method is the possibility to present
the numerical solution in a reliable and simple analytical parameterized
form,  extremely convenient in practical calculations of matrix
elements within the BS formalism and for analytical continuation of the
solution back to Minkowski space.

The   method allows us
to describe, in a covariant way, realistic two-body systems, such
as the deuteron, the positronium and the variety of known
mesons, as bound states of quark-antiquark
pairs and to solve the inhomogeneous BS equation for
scattering states in the continuum.

\section{Acknowledgments}
We thank V.A. Karmanov, G.V. Efimov and T. Frederico for
valuable discussions and M. Mangin-Brinet for sending of results within Light Front
calculating.
 S.M.D. and S.S.S. acknowledge the warm
hospitality at the Elementary Particle Physics group,
University of Rostock, where a bulk of this work was performed.
This work was  supported in part by the Heisenberg -
Landau program of the JINR - FRG collaboration and by the
Deutscher Akademischer Austauschdienst.

\appendix

\section{Partial kernels}
\label{app:kernel}
Here we  present the explicit form of the kernels
 $V^{\beta\alpha}_{k'k}(\tilde p)$ which determine  the
 partial kernels $A^{\beta\alpha}_{jl}(\tilde p)$
and $B^{\beta\alpha}_{km}(\tilde p)$, eqs.  (\ref{set1})-(\ref{set4})
and (\ref{set11})-(\ref{set14}).
For this let us introduce an auxiliary quantity
$S^l_{k'k}(\tilde p)$ defined as
\begin{eqnarray}
S^l_{k^{\prime}k}(\tilde p)&\equiv&\int \limits_0^{\pi} d\chi \sin^2\chi
X_{k^{\prime}l}(\chi)X_{kl}(\chi)\,S(p_4,|\bp|)=  
\int \limits_0^{\pi} d\chi  \frac
{\sin^2\chi \, X_{k^{\prime}l}(\chi)X_{kl}(\chi)}{\left (\tilde p^2+m^2-{M^2}/{4}\right)^2+M^2
\tilde p^2 \cos^2\chi} \nonumber\\
&=&
 \sqrt{\frac {2}{\pi}}\frac { l! \ (-2)^{l+1}}{\tilde p M\left (\tilde p^2+m^2-{M^2}/{4}\right) }
\sqrt{\frac{(k^{\prime}+1)(k^{\prime}-l)!}{(k^{\prime}+l+1)!}
\frac{(k+1)(k-l)!}{(k+l+1)!}}\nonumber\\
&\times& C^{l+1}_{\mathrm{min}-l}(iz) (z^2+1)^{\frac {2l+1}{4}}
Q^{l+\frac12}_{\mathrm{max}+\frac 12}(iz), \label{analit}
\end{eqnarray}
where max (min) is the maximum (minimum) index of
$k,k^{\prime}$, $C^{l+1}_{\mathrm{min}-l}(iz)$ and
$Q^{l+\frac12}_{\mathrm{max}+\frac 12}(iz)$  stand for the   Gegenbauer polynomials
and Legendre functions of the second kind, respectively,
of imaginary argument $iz$ with
\begin{eqnarray}
z=\frac{ \tilde p^2+m^2-{M^2}/{4} }{\tilde p M}. \nonumber
\end{eqnarray}
Note that, as follows from  properties of
Gegenbauer polynomials, the quantities $S^l_{k^{\prime}k}$ are different from zero
 for  nonnegative $k,k'$ and $k+k'$  integer. The
 Legendre functions   in eq. (\ref{analit})
for  $l=0,1,2$ explicitly read as
\begin{eqnarray}
&&Q^{\frac 12}_{n+\frac 12}(iz)={\rm e}^{i\pi(2n+3)/4}
\sqrt{\frac{\pi}{2}}(z^2+1)^{-\frac14}
(z-\sqrt{z^2+1})^{n+1}, \nonumber \\
&&
Q^{\frac 32}_{n+\frac 32}(iz)=\left [(n+1)zQ^{\frac 12}_{n+\frac 32}(iz)-i
 (n+2)Q^{\frac 12}_{n+\frac 12}(iz)\right]\frac {1}{\sqrt{z^2+1}},\nonumber\\
\nonumber
&&
Q^{\frac 52}_{n+\frac 52}(iz)= \left[(n+1)zQ^{\frac 32}_{n+\frac 52}(iz)-i
(n+4)Q^{\frac 32}_{n+\frac 32}(iz)\right]\frac {1}{\sqrt{z^2+1}}.
\nonumber\end{eqnarray}
By using the recurrent relations
for the Gegenbauer polynomials \cite{batman},
 the partial kernels $V^{\beta\alpha}_{k'k}$ can be  expressed
  via $S^l_{k^{\prime}k}$ as
\begin{eqnarray}
V^{13}_{kj}&\equiv&\int d\chi \sin ^2 \chi \frac {X_{k0}(\chi) X_{j1}(\chi)}
{A^2+B^2\cos^2\chi}\sin \chi=
\frac 12 \sqrt{\frac {j+2}{j}} S^0_{k,j-1} -\frac 12
\sqrt{\frac {j}{j+2}}S^0_{k,j+1},\nonumber \\
V^{15}_{kj}&\equiv&\int d\chi \sin ^2 \chi \frac {X_{k0}(\chi) X_{j2}(\chi)}
{A^2+B^2\cos^2\chi} \sin^2 \chi=
\frac 14 \sqrt{\frac {(j+2)(j+3)}{(j-1)j}} S^0_{k,j-2}\nonumber\\
 &-&\frac 14 \frac {\sqrt{(j-1)(j+3)}}{(j+1)} \left(\sqrt {\frac{j+2}{j}}+
 \sqrt {\frac {j}{j+2}}\right) S^0_{k,j}
 +\frac 14 \sqrt{\frac{(j-1)j}{(j+2)(j+3)}}S^0_{k,j+2},
\nonumber\\
V^{22}_{kj}&\equiv&\int d\chi \sin ^2 \chi \frac {X_{k0}(\chi) X_{j0}(\chi)}
{A^2+B^2\cos^2\chi} \cos^2 \chi=
\frac 14 (S^0_{k-1,j-1}+S^0_{k-1,j+1}+S^0_{k+1,j-1}+S^0_{k+1,j+1}),
\nonumber \\
V^{24}_{kj}&\equiv&\int d\chi \sin ^2 \chi \frac {X_{k1}(\chi) X_{j0}(\chi)}
{A^2+B^2\cos^2\chi} \cos \chi \sin \chi \nonumber \\
&=&\frac 14 \left( \sqrt {\frac {k+2}{k}}S^0_{k-1,j+1}-
\sqrt {\frac {k}{k+2}}S^0_{k+1,j+1}+\sqrt {\frac {k+2}{k}}S^0_{k-1,j-1}
-\sqrt {\frac {k}{k+2}}S^0_{k+1,j-1}
 \right),
\nonumber\\
V^{34}_{kj}&\equiv&\int d\chi \sin ^2 \chi \frac {X_{k1}(\chi) X_{j1}(\chi)}
{A^2+B^2\cos^2\chi}\cos \chi =
\frac 12 \left(c_1(k) S^1_{k+1,j}+c_2(k) S^1_{k-1,j}
 \right),
\nonumber \\
V^{35}_{kj}&\equiv&\int d\chi \sin ^2 \chi \frac {X_{k1}(\chi) X_{j2}(\chi)}
{A^2+B^2\cos^2\chi} \sin \chi\nonumber \\
&=&\frac 12 \sqrt{\frac {(j+2)(j+3)}{j(j+1)}} S^1_{k,j-1}
 -\frac 12 \sqrt{\frac {(j-1)j}{(j+1)(j+2)}}S^1_{k,j+1},
\nonumber \\
V^{44}_{kj}&\equiv&\int d\chi \sin ^2 \chi \frac {X_{k1}(\chi) X_{j1}(\chi)}
{A^2+B^2\cos^2\chi} \cos^2 \chi= \frac 14 \left( c_1(k) c_1(j)
S^1_{k+1,j+1}+c_1(k) c_2(j)S^1_{k+1,j-1} \right.\nonumber \\
&+&\left. c_2(k)c_1(j) S^1_{k-1,j+1}+c_2(k) c_2(j)S^1_{k-1,j-1}\right),
\nonumber\\
V^{55}_{kj}&\equiv&\int d\chi \sin ^2 \chi \frac {X_{k2}(\chi) X_{j2}(\chi)}
{A^2+B^2\cos^2\chi} \cos^2 \chi=
\frac 14 \left( d_1(k) d_1(j) S^2_{k+1,j+1}+d_1(k) d_2(j)S^2_{k+1,j-1}
\right.\nonumber\\
&+&\left. d_2(k)d_1(j) S^2_{k-1,j+1}+d_2(k) d_2(j)S^2_{k-1,j-1}\right),
\nonumber\\
V^{57}_{kj}&\equiv&\int d\chi \sin ^2 \chi \frac {X_{k1}(\chi) X_{j2}(\chi)}
{A^2+B^2\cos^2\chi} \cos \chi \sin \chi \nonumber\\
&=&\frac 14  c_1(k) \left(\sqrt {\frac {(j+2)(j+3)}{j(j+1)}}S^1_{k+1,j-1}-
 \sqrt {\frac {(j-1)j}{(j+1)(j+2)}}S^1_{k+1,j+1} \right)\nonumber\\
&+&\frac 14  c_2(k) \left(\sqrt {\frac {(j+2)(j+3)}{j(j+1)}}S^1_{k-1,j-1}-
 \sqrt {\frac {(j-1)j}{(j+1)(j+2)}}S^1_{k-1,j+1} \right),
\nonumber \end{eqnarray}
\begin{eqnarray}
c_1(k)=\sqrt{\frac {k(k+3)}{(k+1)(k+2)}}, \quad \quad
c_2(k)=\sqrt{\frac {(k-1)(k+2)}{k(k+1)}},
\nonumber
\end{eqnarray}
\begin{eqnarray}
&&
A\equiv \tilde p^2 +m^2-M^2/4,\qquad
B\equiv\tilde p M, \nonumber\\&&
d_1(k)=\sqrt{\frac {(k-1)(k+4)}{(k+1)(k+2)}}, \qquad
d_2(k)=\sqrt{\frac {(k-2)(k+3)}{k(k+1)}}.
\nonumber
\end{eqnarray}

\newpage

\begin{table}[h]    
\[
\begin{array}{|c|cccc|}
\hline
$$^1S_0$$& $$^1S_0^{++}$$ & $$^1S_0^{--}$$ & $$^3P_0^e$$ & $$^3P_0^o$$ \\
\hline\hline
$$^3S_1$--$^3D_1$$ &$$^3S_1^{++}$$ &$$^3S_1^{--}$$ &
$$^3D_1^{++}$$ & $$^3D_1^{--}$$  \\
&$$^3P_1^{e}$$ &$$^3P_1^{o}$$ & $$^1P_1^{e}$$ & $$^1P_1^{o}$$  \\
 \hline
\end{array}
\]\caption{Classification  of the $\rho$-spin
partial components for the $^1S_0$ and $^3S_1$--$^3D_1$ channels
in spectroscopic notation.}
\label{tab1}
\end{table}


\begin{table}[h]  
\[
\begin{array}{|c||c|r|r|}
\hline
n & b_n[S] & b_n[PS] & b_n[V]\\ \hline \hline
1 & 1 & - 1 & 4\\ \hline
2 & 1 & 1 & -2\\ \hline
3 & 1 & - 1 &  0\\ \hline
4 & 1 & 1 & -2\\  \hline
\end{array}
\]\caption{The coefficients $b_n$, eq. (\ref{par1s0}),
in the $^1S_0$ channel with
scalar (S), pseudoscalar (PS) and vector (V, without the tensor
part) meson exchanges.}
\label{tab5}
\end{table}

\begin{table}[h]     
\[
\begin{array}{|c||c|r|r|}
\hline
n & b_n[S] & b_n[PS] & b_n[V]\\
\hline
\hline 1 & 1 & 1 & 2\\
\hline 2 & 1 & -1 & 0\\
\hline 3 & 1 & 1 & -4\\
\hline 4 & 1 & -1 & -2\\
\hline 5 & 1 & 1 & 2\\
\hline 6 & 1 & -1 & 0\\
\hline 7 & 1 & 1 & 2\\
\hline 8 & 1 & -1 & 0\\
 \hline
\end{array}
\]\caption{The same as in Table~\ref{tab5} but for the
$^3S_1$--$^3D_1$ channel.}
\label{tab6}
\end{table}
\newpage

\begin{table}[h]  
\[
\begin{array}{|c||c|c|c|c|}
\hline
n & a_{n 1} & a_{n 2} & a_{n 3} & a_{n 4}\\\hline\hline
1 & \frac{M^2}{4}+m^2+\tilde k^2& mM & -M|\bk| & 0 \\\hline
2 & mM & \frac{M^2}{4}m^2+2k_4^2-\tilde k^2 & -2m |\bk| &-2k_4 |\bk| \\\hline
3 & M |\bk| & 2m|\bk| & -\frac{M^2}{4}+m^2-\tilde k^2 & 2mk_4 \\\hline
4 & 0 & -2k_4 |\bk| & -2mk_4 & -\frac{M^2}{4}+m^2-2k_4^2+\tilde k^2 \\ \hline
\end{array}
\]\caption{The quantities $a_{nm}$, eq. (\ref{par2s0}), for the $^1S_0$ state.}
\label{tab2}
\end{table}
\newpage

\begin{table}[h]    
\[
\begin{array}{|c||c|c|c|c|} \hline
n & a_{n 1} & a_{n 2} & a_{n 3} & a_{n 4}\\\hline\hline
1 & \frac{M^2}{4}+m^2+\frac23 k_4^2+\frac{\tilde k^2}{3} & mM &
-\frac{2\sqrt{3}}{3}m|\bk| &- \frac{\sqrt{6}}{3}M|\bk| \\\hline
2 & mM & \frac{M^2}{4}+m^2+\frac43 k_4^2-\frac{\tilde k^2}{3} &
-\frac{\sqrt{3}}{3}M|\bk| &- \frac{2\sqrt{6}}{3}m|\bk| \\\hline
3 & \frac{2\sqrt{3}}{3}m|\bk|  & \frac{\sqrt{3}}{3}M|\bk| &
-\frac{M^2}{4}+m^2- \tilde k^2 & 0 \\\hline
4 & \frac{\sqrt{6}}{3}M|\bk| & \frac{2\sqrt{6}}{3}m|\bk| &
0 & -\frac{M^2}{4}+m^2-\tilde k^2 \\ \hline
5 & \frac{2\sqrt{2}}{3}|\bk|^2 & 0 & \frac{2\sqrt{6}}{3}m|\bk| &
-\frac{\sqrt{3}}{3}M|\bk|\\ \hline
6 & 0 & -\frac{2\sqrt{2}}{3}|\bk|^2 & \frac{\sqrt{6}}{3}M|\bk| &
-\frac{2\sqrt{3}}{3}m|\bk|\\ \hline
7 & \frac{2\sqrt{3}}{3}k_4 |\bk|& 0 & 2mk_4 & 0 \\ \hline
8 & 0 & \frac{2\sqrt{6}}{3}k_4|\bk| & 0 & 2mk_4 \\ \hline
\end{array}
\]\caption{The quantities $a_{nm}$,
eq.~(\ref{par2s0}), for the $^3S_1$--$^3D_1$ state.}
\label{tab3}
\end{table}
\newpage

\begin{table}[ht]     
\[
\begin{array}{|c||c|c|c|c|}\hline
n  & a_{n 5} & a_{n 6} & a_{n 7} & a_{n 8}\\\hline\hline
1  & \frac{2\sqrt{2}}{3}|\bk|^2      & 0 & \frac{2\sqrt{3}}{3}k_4|\bk| &0 \\ \hline
2  & 0 & -\frac{2\sqrt{2}}{3}|\bk|^2 & 0 & \frac{2\sqrt{6}}{3}k_4|\bk|\\ \hline
3  & -\frac{2\sqrt{6}}{3}m|\bk|& -\frac{\sqrt{6}}{3}M|\bk| & -2mk_4 & 0 \\ \hline
4  & \frac{\sqrt{3}}{3}M|\bk| & \frac{2\sqrt{3}}{3}m|\bk| & 0 & -2mk_4 \\ \hline
5  & \frac{M^2}{4}+m^2+\frac43 k_4^2-\frac{\tilde k^2}{3}& mM &-\frac{2\sqrt{6}}{3}k_4|\bk| & 0 \\\hline
6  & mM & \frac{M^2}{4}+m^2+\frac 23 k_4^2+\frac{\tilde k^2}{3} &0 & \frac{2\sqrt{3}}{3}k_4|\bk| \\\hline
7  & -\frac{2\sqrt{6}}{3}k_4|\bk| & 0 &-\frac{M^2}{4}+m^2-2k_4^2+\tilde k^2 & 0 \\\hline
8  & 0 & \frac{2\sqrt{3}}{3}k_4|\bk| &0 & -\frac{M^2}{4}+m^2-2k_4^2+\tilde k^2 \\\hline
\end{array}
\]
\caption{Continuation of Table \ref{tab3}.}
\label{tab4}
\end{table}
\newpage

\begin{table}[h]     
\[
\begin{array}{|c||c|c|c|c|}
\hline
n & A^{n 1}_{jj'} & A^{n 2}_{jj'} & A^{n 3}_{jj'} & A^{n 4}_{jj'}\\
\hline
\hline
1 & -(M^2/4+m^2+{\tilde k}^2)& -mM  & {\tilde k}M & 0 \\
& \times S^0_{2j-2,2j'-2} &\times S^0_{2j-2,2j'-2}& \times V^{13}_{2j-2,2j'-1}&\\\hline
 &  & -(M^2/4+m^2-{\tilde k}^2) &  &  \\
2 & -mM\, S^0_{2j-2,2j'-2}&\times S^0_{2j-2,2j'-2}
&2m{\tilde k}\, V^{13}_{2j-2,2j'-1}
& 2{\tilde k}^2\, V^{24}_{2j',2j-2}\\
& &-2{\tilde k}^2V^{22}_{2j-2,2j'-2}& &\\
\hline
3 & -{\tilde k}M & -2m{\tilde k} & (M^2/4-m^2+{\tilde k}^2) & -2m{\tilde k} \\
& \times V^{13}_{2j'-2,2j-1} & \times V^{13}_{2j'-2,2j-1} &
\times S^1_{2j-1,2j'-1} & \times V^{34}_{2j-1,2j'}\\
\hline
4 & 0 & 2{\tilde k}^2\, V^{24}_{2j,2j'-2} & 2j'{\tilde k}\, V^{34}_{2j,2j'-1}
& (M^2/4-m^2-{\tilde k}^2) \\
 &  & &  &\times S^1_{2j,2j'}+2{\tilde k}^2V^{44}_{2j',2m}\\
 \hline
\end{array}
\]\caption{The partial kernels
$A^{n n'}_{jj'}(\tilde k)$ defined by  eqs. (\ref{set1})-(\ref{set4}).
For the explicit form of
the introduced quantities $S^{0,1}_{k'k}(\tilde k)$ and
$V^{n n'}_{j j'}(\tilde k)$ see Appendix \ref{app:kernel}.}\label{tab7}
\end{table}

\newpage

\begin{table}
\[
\begin{array}{|c||c|c|c|c|}
\hline
n & B^{n 1}_{jj'} & B^{n 2}_{jj'} & B^{n 3}_{jj'} & B^{n 4}_{jj'}\\ \hline\hline
 & -(M^2/4+m^2+{\tilde k}^2/3)&   & 2m{\tilde k} & \sqrt{2}M{\tilde k} \\
1& \times S^0_{2j-2,2j'-2} &-mM S^0_{2j-2,2j'-2}&
\times V^{13}_{2j-2,2j'-1}/\sqrt{3}& \times V^{13}_{2j-2,2j'-1}/\sqrt{3}\\
&-2{\tilde k}^2V^{22}_{2j-2,2j'-2}/3 & & & \\ \hline
 &  & -(M^2/4+m^2-{\tilde k}^2/3) & M{\tilde k} & 2\sqrt{6}\,m{\tilde k} \\
2 & -mM\, S^0_{2j-2,2j'-2}&\times S^0_{2j-2,2j'-2}
&\times V^{13}_{2j-2,2j'-1}/\sqrt{3}
& \times V^{13}_{2j-2,2j'-1}/3\\
& &-4{\tilde k}^2V^{22}_{2j-2,2j'-2}/3& &\\ \hline
3 & -2m{\tilde k} & -{\tilde k}M & (M^2/4-m^2+{\tilde k}^2) & 0 \\
& \times V^{13}_{2j'-2,2j-1}/\sqrt{3} & \times V^{13}_{2j'-2,2j-1}/\sqrt{3} &
\times S^1_{2j-1,2j'-1} & \\\hline
4 & -\sqrt{6}\,{\tilde k}M & -2\sqrt{6}\,m{\tilde k} & 0 & (M^2/4-m^2+{\tilde k}^2) \\
 & \times V^{13}_{2j'-2,2j-1}/3 & \times V^{13}_{2j'-2,2j-1}/3
&  & \times S^1_{2j-1,2j'-1}\\ \hline
5& -2\sqrt{2}\,{\tilde k}^2V^{15}_{2j'-2,2j}/3 & 0
& -2\sqrt{6}\,m{\tilde k}& {\tilde k}MV^{35}_{2j'-1,2j}/\sqrt{3}\\
& & & \times V^{35}_{2j'-1,2j}/3 & \\  \hline
6& 0 & 2\sqrt{2}\,{\tilde k}^2V^{15}_{2j'-2,2j}/3 & -\sqrt{6}\,{\tilde k}M
& 2m{\tilde k}V^{35}_{2j'-1,2j}/\sqrt{3}\\
& & & \times V^{35}_{2j'-1,2j}/3 & \\ \hline
7& -2{\tilde k}^2V^{24}_{2j,2j'-2}/\sqrt{3} & 0 & -2m{\tilde k}V^{34}_{2j,2j'-1} & 0 \\ \hline
8& 0 & -2\sqrt{6}\,{\tilde k}^2V^{24}_{2j,2j'-2}/3 & 0 & -2m{\tilde k}V^{34}_{2j,2j'-1}\\
\hline
\end{array}
\]\caption{The partial kernels
$B^{nn' }_{jj'}(\tilde k)$ defined by eqs.  (\ref{set11})-(\ref{set14}).
For the explicit form of the quantities $S^{0,1}_{jj'}(\tilde k)$ and
$V^{nn'}_{jj'}(\tilde k)$ see Appendix \ref{app:kernel}.}\label{tab8}
\end{table}
\newpage

\begin{table}
\[
\begin{array}{|c||c|c|c|c|}
\hline
n & B^{n 5}_{jj'} & B^{n 6}_{jj'} & B^{n 7}_{jj'} & B^{n 8}_{jj'}\\ \hline\hline
1& -2\sqrt{2}\,{\tilde k}^2V^{15}_{2k-2,2m}/3 & 0 & -2{\tilde k}^2& 0\\
& & & \times V^{24}_{2j',2j-2}/\sqrt{3} & \\ \hline
2& 0 & 2\sqrt{2}\,{\tilde k}^2V^{15}_{2j'-2,2j}/3 & 0
& -2\sqrt{6}\,{\tilde k}^2\\
& & &  & \times V^{24}_{2j,2j'-2}/3\\
\hline
3& 2\sqrt{6}\,m{\tilde k}V^{35}_{2j-1,2j'}/3 & \sqrt{6}\,{\tilde k}MV^{35}_{2j-1,2j'}/3
& 2m{\tilde k}V^{34}_{2j-1,2j'} & 0 \\ \hline
4& -{\tilde k}MV^{35}_{2j-1,2j'}/\sqrt{3}
& -2m{\tilde k}V^{35}_{2j-1,2j'}/\sqrt{3} & 0 & 2m{\tilde k}V^{34}_{2j-1,2j'}\\ \hline
5 & -(M^2/4+m^2-{\tilde k}^2/3)& -mM S^2_{2j,2j'}  & 2\sqrt{6}\,{\tilde k}^2 & 0 \\
& \times S^2_{2j,2j'}-4{\tilde k}^2V^{55}_{2j,2j'}/3 & &
\times V^{57}_{2j',2j}/3& \\ \hline
6 & -mM\, S^2_{2j,2j'} & -(M^2/4+m^2+{\tilde k}^2/3) & 0 & -2{\tilde k}^2 \\
 & &\times S^2_{2j,2j'}-2{\tilde k}^2V^{55}_{2j,2j'}/3
& & \times V^{57}_{2j',2j}/\sqrt{3}\\ \hline
 &  &  & (M^2/4-m^2-{\tilde k}^2) &  \\
7 & 2\sqrt{6}\,{\tilde k}^2V^{57}_{2j,2j'}/3 & 0 &
\times S^1_{2j,2j'}& 0 \\
&  &  & +2{\tilde k}^2V^{44}_{2j,2j'} & \\ \hline
 &  &  &  & (M^2/4-m^2-{\tilde k}^2) \\
8 & 0 & -2{\tilde k}^2 V^{57}_{2j,2j'}/\sqrt{3}
& 0 & \times S^1_{2j,2j'}\\
 &  &  &  & +2{\tilde k}^2V^{44}_{2j,2j'} \\ \hline
\end{array}
\]\caption{Continuation of Table \ref{tab8}.}\label{tab9}
\end{table}

\newpage

\begin{table}
\[
\begin{array}{|c||c|c|c||c|c|c|}
\hline g^2=15 & \multicolumn{3}{c||}{\mu=0.15~{GeV/c^2}}
 & \multicolumn{3}{c|}{\mu=0.5~ {GeV/c^2}} \\
\hline M_\mathrm{max}  & N_G=32 & N_G=64 & N_G=96 & N_G=32 & N_G=64 & N_G=96 \\
\hline
\hline 1 & 1.9399 & 1.9399 & 1.9399 & 1.9984 & 1.9984 & 1.9984\\
\hline 2 & 1.9370 & 1.9370 & 1.9370 & 1.9982 & 1.9982 & 1.9982\\
\hline 3 & 1.9368 & 1.9368 & 1.9368 & 1.9982 & 1.9982 & 1.9982\\
\hline 4 & 1.9368 & 1.9368 & 1.9368 & 1.9982 & 1.9982 & 1.9982\\
\hline \multicolumn{7}{c}{}\\
\hline g^2=30 & \multicolumn{3}{c||}{\mu=0.15~ {GeV/c^2}}
 & \multicolumn{3}{c|}{\mu=0.5~ {GeV/c^2}} \\
\hline M_\mathrm{max}  & N_G=32 & N_G=64 & N_G=96 & N_G=32 & N_G=64 & N_G=96 \\
\hline
\hline 1 & 1.7932 & 1.7910 & 1.7905 & 1.9167 & 1.9142 & 1.9137\\
\hline 2 & 1.7897 & 1.7875 & 1.7871 & 1.9152 & 1.9127 & 1.9122\\
\hline 3 & 1.7896 & 1.7874 & 1.7870 & 1.9152 & 1.9127 & 1.9122\\
\hline 4 & 1.7896 & 1.7874 & 1.7870 & 1.9152 & 1.9127 & 1.9122\\
\hline
\end{array}
\]\caption{Convergence of the calculated values of mass $M$
with respect to the number of points in Gaussian mesh $N_G$ and
the number of hyperspherical components $M_\mathrm{max}$ for the values
of the coupling constant $g^2=15$ and $g^2=30$ in $^1S_0$
channel for the scalar meson exchange kernel.}\label{tab10}
\end{table}
\newpage

\begin{table}
\[
\begin{array}{|c||c|c|c|c|c|c|}
\hline j & b_j & a^j_1 & a^j_2 & a^j_3 & a^j_4 & \chi^2
\\
\hline\hline
1& 0.8162 & 3.7476 & -3.3910 & 2.9564 & -0.0107 & 3.4\cdot 10^{-4}\\[1ex]
2& 0.5875 & -3.9733 & 6.6483 & -6.5271 & 3.6482 & 1.5\cdot 10^{-6}\\[1ex]
3& 0.4556 & 4.2805 & -5.6270 & 6.1247 & -4.8948 & 2.4\cdot 10^{-6}\\[1ex]
4& 0.3680 & -4.2026 & 9.4868 & -22.191 & 17.300 & 8\cdot 10^{-5}\\[1ex]
 \hline
\end{array}
\]\caption{Numerical values of the parameters in formula (\ref{fit}) and the corresponding
$\chi^2$ values.
}\label{tab11}
\end{table}

\newpage

\begin{table}[t]                   
\[
\begin{array}{|c|c|c|c|c|c|}
\hline g^2 & M, GeV/c^2 & P_{++}  & P_{--} & P_{o}  & P_{e}
\\
\hline\hline
15 & 1.937 & 1.012 & -1.18\cdot 10^{-3} & -6.63\cdot 10^{-3} & -4.37\cdot 10^{-3}\\[1ex]
20 & 1.892 & 1.020 & -2.99\cdot 10^{-3} & -1.07\cdot 10^{-2} & -6.92\cdot 10^{-3}\\[1ex]
25& 1.842 & 1.030 & -6.22\cdot 10^{-3} & -1.46\cdot 10^{-2} & -9.41\cdot 10^{-3}\\[1ex]
27 & 1.820 & 1.034 & -8.11\cdot 10^{-3} & -1.61\cdot 10^{-2} & -1.03\cdot 10^{-2}\\[1ex]
29 & 1.798 & 1.039 & -1.05\cdot 10^{-2} & -1.75\cdot 10^{-2} & -1.12\cdot 10^{-2}\\[1ex]
29.5 & 1.788 & 1.041 & -1.25\cdot 10^{-2} & -1.80\cdot 10^{-2} & -1.16\cdot 10^{-2}\\[1ex]
29.6 & 1.5 & 1.210 & -0.19            & -1.24\cdot 10^{-2} & -7.77\cdot 10^{-3}\\[1ex]
 \hline
\end{array}
\]
\caption{Pseudo-probabilities, eq. (\ref{pseudopr}),
  in the $^1S_0$ state at given $g^2$ and $M$,
  normalized as  (\ref{normir}).
 In the LF formalism only
$''++''$ and $''o''$ components are relevant to the
computed pseudo-probabilities.}
\label{tab13}
\end{table}

\newpage

\begin{table}
\[
\begin{array}{|c||c|c|c|c|c|c|c|c|}
\hline\Lambda,m_ec^2 &10 & 20 & 30 & 40 & 50 & 60 & 70 &80
\\
\hline
B_{^1S_0}\cdot 10^2&1.378& 1.362 & 1.360 & 1.355 & 1.352 & 1.350 & 1.349& 1.348\\[1ex]
B_{^3S_1}\cdot 10^2&1.273& 1.268 & 1.262 & 1.260 & 1.259 & 1.258 & 1.257& 1.256\\
 \hline
\end{array}
\]\caption{Binding energies of positronium  states  as  a function  of a cutoff parameter  $\Lambda$
in units of $m_e c^2$ at $g^2=3.77$.
}\label{tab12}
\end{table}

\newpage

\newpage

\newpage

\newpage

\newpage

\newpage

\newpage

\newpage

\newpage

\end{document}